# THERMODYNAMICS OF MIXTURES CONTAINING AROMATIC NITRILES

JUAN ANTONIO GONZÁLEZ*[1], CRISTINA ALONSO TRISTÁN[2], FERNANDO HEVIA,[1] , ISAÍAS GARCÍA DE LA FUENTE[1] AND LUIS FELIPE SANZ[1]

[1] Dpto. Ingeniería Electromecánica. Escuela Politécnica Superior. Avda. Cantabria s/n. 09006 Burgos, (Spain)

[2] G.E.T.E.F., Departamento de Física Aplicada, Facultad de Ciencias, Universidad de Valladolid, Paseo de Belén, 7, 47011 Valladolid, Spain,

*e-mail: jagl@termo.uva.es; Fax: +34-983-423136; Tel: +34-983-423757

**ABSTRACT**

The coexistence curves of liquid-liquid equilibrium (LLE) for the mixtures: phenylacetonitrile + heptane, + octane, + nonane, + cyclooctane, or + 2,2,4-trimethylpentane and for 3-phenylpropionitrile + heptane, or + octane are reported. Aromatic nitrile + alkane, + aromatic hydrocarbon or + 1 alkanol systems are investigated using a set of thermophysical properties: phase equilibria (solid-liquid, SLE, vapour-liquid, VLE and LLE), excess molar functions, enthalpies ($H_{\mathrm{m}}^{\mathrm{E}}$), isochoric internal energies ($U_{\mathrm{Vm}}^{\mathrm{E}}$), isobaric heat capacities ($C_{\mathrm{pm}}^{\mathrm{E}}$) and volumes ($V_{\mathrm{m}}^{\mathrm{E}}$), and the Kirkwood's correlation factor. Due to proximity effects between the phenyl and the CN groups, dipolar interactions between molecules of aromatic nitriles are stronger than those between molecules of isomeric linear nitriles. Dipolar interactions become weaker in the order: 3-phenylpropionitrile > phenylacetonitrile > benzonitrile. Benzonitrile + aromatic hydrocarbon mixtures are characterized by dispersive interactions and structural effects. The latter are more important in systems with phenylacetonitrile. Structural effects are also present in benzonitrile + *n*-alkane, or + 1-alkanol + mixtures. The systems mentioned above have been studied using DISQUAC. Interaction parameters for contacts where the CN group in aromatic nitriles participates are given. DISQUAC describes correctly any type of phase equilibria, $C_{\mathrm{pm}}^{\mathrm{E}}$ of benzonitrile + hydrocarbon mixtures and $H_{\mathrm{m}}^{\mathrm{E}}$ of benzonitrile + cyclohexane, or 1-alkanol systems. Large differences encountered between theoretical $H_{\mathrm{m}}^{\mathrm{E}}$ values and experimental data for some solutions are discussed. 1-Alkanol + benzonitrile mixtures are also investigated by means of the ERAS model. ERAS represents well $H_{\mathrm{m}}^{\mathrm{E}}$ of these systems. The $V_{\mathrm{m}}^{\mathrm{E}}$ curves of solutions with longer 1-alkanols are more poorly described, which has been explained in terms of the existence of structural effects.

**KEYWORDS: LLE; aromatic nitrile; proximity effects; dipolar interactions; DISQUAC**

## 1. Introduction

Aromatic nitriles, $C_6H_5-(CH_2)_{n-1}CN$ ($n$ = 1 (benzonitrile); $n$ = 2 (phenylacetonitrile); $n$ = 3 (3-phenylpropionitrile)) are rather polar compounds as it is indicated by their high dipole moments ($\mu$) [1]: 3.87 D ($n$ =1); 3.50 D ($n$ =2); 3.29 D ($n$ =3). They are good solvents and are used as starting materials in the synthesis of fungicides, fragances and pharmaceuticals, as analgesics or antihistamines. Recently, the mixture formed by 3-phenylpropionitrile and supercritical $CO_2$ has received attention as the miscibility of these type of systems is a relevant condition for polymer processes [2].

It is well known that the existence of two groups (X,Y) of the same or different nature within the same molecule leads to an enhancement of the dipolar interactions between the mentioned molecules. For example, the 1-butanol + heptane mixture is miscible at any concentration at 298.15 K and, at equimolar composition, its excess molar enthalpy, $H_m^E$, is 575 J·mol$^{-1}$ [3]. In contrast, the 2-methoxyethanol (isomeric molecule of 1-butanol) + heptane system is characterized by a moderately high UCST (319.74 K [4]), what has been typically ascribed to proximity effects between the $-O-$ and $-OH$ groups within the alkoxyethanol [5]. It is also known, that the application of group contribution methods to systems characterized by proximity effects between two given groups X,Y leads to erroneous results when one uses interaction parameters for the (X/Y) contacts obtained from solutions with X and Y belonging to different molecules [6]. In fact, proximity effects can drastically change the interaction parameters of the applied model. Thus, in terms of UNIFAC (Dortmund version) specific main groups have been defined for aniline or phenol for a better representation of the thermodynamic properties of mixtures involving these compounds [7]. We are engaged in a systematic research on proximity effects between the phenyl group $C_6H_5-$ and a polar group as carbonyl, aldehyde, alkanoate, amine, alkanol or oxygen. At this end, we have provided LLE measurements for systems formed by one alkane and a polar aromatic compound such as: $C_6H_5-(CH_2)_{n-1}COCH_3$ ($n$ =1,2,3) [8-10]; $C_6H_5-CHO$ [11]; $C_6H_5-CH_2COOCH_3$ [9,10] $C_6H_5-CH_2NH_2$ [12]; $C_6H_5-CH_2OH$ [13], or $C_6H_5-O-CH_2CH_2OH$ [14]. These data together with measurements available in the literature on VLE or $H_m^E$ for these or related compounds, aniline or phenol, e.g., have been employed for the characterization of aromatic polar compound + organic solvent mixtures [8-16] in terms of the DISQUAC group contribution model [17]. Here, we continue with this research line and report LLE data for the systems: $C_6H_5-CH_2CN$ + heptane, or + octane, or + nonane, or + cyclooctane, or + 2,2,4-trimethylpentane, and for $C_6H_5-CH_2CH_2CN$ + heptane, or + octane. In addition, aromatic nitrile + organic solvent mixtures are characterized using DISQUAC and the interaction parameters for a number of contacts, CN/aliphatic; CN/aromatic and CN/OH, are provided. UNIFAC (Dortmund) and DISQUAC interaction parameters for the contacts CN/aliphatic and CN/OH in systems involving linear nitriles are available in the literature [7,18,19]. The mixtures 1-alkanol + linear nitrile or + benzonitrile have been investigated by means of the Flory theory [20], and the ERAS model [21] has been applied to 1-alkanol + benzonitrile systems [22]. On the other hand, due to high $\mu$ values of aromatic nitriles, one may expect the existence of strong dipolar interactions between nitrile molecules in solutions where these compounds participate. An interesting question arises from the concentration dependence of $H_m^E$ of benzonitrile + benzene, or + toluene systems. Some authors have reported M-shaped $H_m^E$ curves for these mixtures [23,24]. Such anomalous concentration dependence of $H_m^E$ has been explained by invoking charge-transfer complex formation between benzonitrile and the aromatic hydrocarbon [23]. However, it must be remarked that other researchers have not observed the mentioned trends for benzonitrile mixtures [25]. This matter will also be considered along the work.

## 2. Experimental

### *2.1 Materials.*

Table 1 shows some properties of the pure chemicals used along the experimental part of this investigation: source, purity, water contents, determined by the Karl-Fischer method, and density ($\rho$). The reagents were employed without further purification. Density values were determined by means of a vibrating-tube densimeter and a sound analyser, Anton Paar model DSA-5000. The repeatability of the $\rho$ measurements is $5 \cdot 10^{-3}$ $kg \cdot m^{-3}$, while their relative standard uncertainty is 0.0012. From the values collected in Table 1, we can conclude that there is a good agreement between our density results and those reported in the literature.

### *2.2 Apparatus and Procedure*

Mixtures were prepared by mass in small Pyrex tubes (0.009 m i.d. and about 0.04 m length, with a free volume of the ampoule $\approx 1.17 \cdot 10^{-6}$ $m^3$). The tubes were immediately sealed by capping at 0.1 MPa and 298.15 K. Weights were determined by means of an analytical balance HR-202 (weighing accuracy $10^{-8}$ kg). Mole fractions were calculated using the relative atomic mass Table of 2015 issued by the Commission on Isotopic Abundances and Atomic Weights (IUPAC) [26]

As in other applications, the coexistence curves of liquid-liquid equilibrium were determined by the method of the critical opalescence. Details on the experimental technique have been previously reported [14]. The equilibrium temperatures were measured using a Pt-1000 resistance, calibrated, according to the ITS-90 scale of temperature, against the triple point of the water and the fusion point of Ga. The precision of the equilibrium temperature measurements is ± 0.001 K. The estimated standard uncertainty is 0.1 K in the flat region of the coexistence curves. The standard uncertainty of the equilibrium mole fraction is 0.0005. This value of composition uncertainty takes into account that the more volatile component is partially evaporated to the mentioned free volume of the ampoule.

## 3. Experimental results

Table 2 lists the directly measured liquid-liquid equilibrium temperatures, $T$, vs. $x_1$, the mole fraction of the aromatic nitrile, for the systems: phenylacetonitrile + *n*-$C_7$, or + *n*-$C_8$, or + *n*-$C_9$, or + cyclooctane, or + 2,3,4-trimethylpentane, and for 3-phenylpropionitrile + *n*-$C_7$, or + *n*-$C_8$. A literature survey shows that there are no data available for comparison. Some experimental results are also represented in Figure 1.

For the considered systems, the LLE curves are characterized by a rather flat maximum and become progressively skewed towards higher $x_1$ values when the number of C atoms of the *n*-alkane is increased. In addition, the upper critical solution temperature, UCST, of the studied solutions increases with the *n*-alkane size. These features are also encountered in many others systems previously investigated as those formed by *n*-alkane and acetophenone [8], phenylacetone [9], or benzaldehyde [11], or aromatic alcohols [13], or aromatic amines, or linear organic carbonate, or acetic anhydride, or alkoxyethanol, or linear polyether (see reference [9] for the source of experimental data).

The experimental ($x_1$,$T$) pairs obtained for each system were correlated with the equation [27,28]:

$$T / K = T_c / K + k\left|y - y_c\right|^m \tag{1}$$

where

$$y = \frac{\alpha\, x_1}{1 + x_1(\alpha - 1)} \tag{2}$$

$$y_c = \frac{\alpha\, x_{1c}}{1 + x_{1c}(\alpha - 1)} \tag{3}$$

In equations (1)-(3), $m$, $k$, $\alpha$, $T_c$ and $x_{1c}$ are the parameters which must be adjusted against the experimental data. Particularly, ($x_{1c}$, $T_c$) stand for the coordinates of the critical point. It should be remarked that, when $\alpha$ = 1, equation (1) is similar to [29-31]:

$$\Delta\lambda = B\tau^{\beta} \tag{4}$$

In equation (4), $\Delta\lambda_1 = \lambda_1' - \lambda_2''$ may be any density variable in the conjugate phase (order parameter). In this investigation, $\lambda_1 = x_1$. On the other hand, τ (= $T_c$ – $T$)/$T_c$ ) denotes the reduced temperature and β is the critical exponent linked to $\Delta\lambda_1$. The critical exponent β value depends on the theory applied to its determination [29-32].

The adjustment of the $m$, $k$, α, $T_c$ and $x_{1c}$ parameters was conducted by means of the Marquardt algorithm [33] with all the points weighted equally. Final values of $m$, $k$, α, $T_c$ and $x_{1c}$ and of the standard deviations for LLE temperatures, $\sigma(T)$, are given in Table 3. The corresponding $\sigma(T)$ values are calculated from:

$$\left(\sigma(T)/K\right) = \left[\sum\left(T_{\rm exp}/K - T_{\rm calc}/K\right)^2 / \left(N-n\right)\right]^{1/2} \tag{5}$$

Here, $N$ stands for the number of data points, and $n$ (= 5) is the number of adjusted parameters. Equation (1) fits well the experimental measurements.

## 4. Models

### *4.1 DISQUAC*

DISQUAC is a group contribution model based on the rigid lattice theory developed by Guggenheim [34]. Some important features of DISQUAC are the following. (i) The total molecular volumes, $r_i$, surfaces, $q_i$, and the molecular surface fractions, $\alpha_{si}$, of the compounds present in the mixture are calculated additively on the basis of the group volumes $R_G$ and surfaces $Q_G$ recommended by Bondi [35]. The volume $R_{CH4}$ and surface $Q_{CH4}$ of methane are taken arbitrarily as equal to 1 [36]. The geometrical parameters for the groups referred to in this work are available in the literature [18,36-38] (ii) The partition function is factorized into two terms, and the excess functions are calculated as the sum of two contributions: a dispersive (DIS) term related to the contribution from the dispersive forces; and a quasichemical (QUAC) term due to the anisotropy of the field forces created by the solution molecules. In the case of the Gibbs energy, $G_{\rm m}^{\rm E}$, a combinatorial term, $G_{\rm m}^{\rm E,COMB}$, represented by the Flory-Huggins equation [36,39] must be included. Thus,

$$G_{\rm m}^{\rm E} = G_{\rm m}^{\rm E,DIS} + G_{\rm m}^{\rm E,QUAC} + G_{\rm m}^{\rm E,COMB} \tag{6}$$

$$H_{\rm m}^{\rm E} = H_{\rm m}^{\rm E,DIS} + H_{\rm m}^{\rm E,QUAC} \tag{7}$$

(iii) The interaction parameters are assumed to be dependent on the molecular structure; (iv) The value $z$ = 4 for the coordination number is used for all the polar contacts. This important shortcoming of the model is partially removed via the hypothesis of considering structure dependent interaction parameters. (v) $V_{\mathrm{m}}^{\mathrm{E}}$ (excess molar volume) = 0 is assumed.

The equations used to calculate the DIS and QUAC contributions to $G_{\mathrm{m}}^{\mathrm{E}}$ and $H_{\mathrm{m}}^{\mathrm{E}}$ in the framework of DISQUAC are given elsewhere [40]. The temperature dependence of the interaction parameters is expressed in terms of the DIS and QUAC interchange coefficients [40], $C_{\mathrm{st,l}}^{\mathrm{DIS}}; C_{\mathrm{st,l}}^{\mathrm{QUAC}}$ where s ≠ t are two contact surfaces present in the mixture and $l$ = 1 (Gibbs energy; $C_{\mathrm{st,1}}^{\mathrm{DIS/QUAC}} = g_{\mathrm{st}}^{\mathrm{DIS/QUAC}}(T_{\mathrm{o}}) / RT_{\mathrm{o}}$); $l$ = 2 (enthalpy, $C_{\mathrm{st,2}}^{\mathrm{DIS/QUAC}} = h_{\mathrm{st}}^{\mathrm{DIS/QUAC}}(T_{\mathrm{o}}) / RT_{\mathrm{o}}$)), $l$ = 3 (heat capacity, $C_{\mathrm{st,3}}^{\mathrm{DIS/QUAC}} = c_{\mathrm{pst}}^{\mathrm{DIS/QUAC}}(T_{\mathrm{o}}) / R$)). $T_{\mathrm{o}}$ = 298.15 K is the scaling temperature and $R$, the gas constant. The equations can be found elsewhere [40].

As usually, DISQUAC calculations on LLE were performed taking into account that the values of the mole fraction $x_1$ of component 1 ($x_1^{'}, x_1^{''}$) relating to the two phases in equilibrium are such that the functions $G_{\mathrm{m}}^{\mathrm{M'}}, G_{m}^{M''}$ ($G_{\mathrm{m}}^{\mathrm{M}} = G_{\mathrm{m}}^{\mathrm{E}} + G_{\mathrm{m}}^{\mathrm{ideal}}$) have a common tangent [41].

The equation of the solid-equilibrium curve of a pure solid component 1 in a solvent mixture is [42]:

$$-\ln x_1 = (\Delta H_{m1} / R)\left[1/T - 1/T_{m1}\right] - (\Delta C_{Pm1} / R)\left[\ln(T / T_{m1}) + (T_{m1} / T) - 1\right] \quad (8)$$

In this equation, $x_1$ is the mole fraction and $\gamma_1$ the activity coefficient of component 1 in the solvent mixture, at temperature $T$. Here, we have used DISQUAC to calculate $\gamma_1$. $\Delta H_{\mathrm{m1}}$ $T_{\mathrm{m1}}$ and $\Delta C_{\mathrm{pm1}}$ are respectively, the enthalpy of fusion, the melting temperature and the molar heat capacity change during the melting process of component 1. All the physical constants needed for calculations are listed in Table S1 (supplementary material). On the other hand, equation (8) assumes that: (i) the phase transition takes place between the system temperature and $T_{\mathrm{m1}}$; (ii) the absence of miscibility in solid phase; (iii) $\Delta C_{\mathrm{pm1}}$ does not depend on the temperature.

*4.2 ERAS*

The main features of this model are now summarized. (i) The excess functions are calculated as the sum of two terms, one arising from hydrogen-bonding effects (the chemical contribution, $X_{\mathrm{m,chem}}^{\mathrm{E}}$) and one related to non-polar van der Waals' interactions including free volume effects (physical contribution, $X_{\mathrm{m,phys}}^{\mathrm{E}}$). The resulting expressions for $X_{\mathrm{m}}^{\mathrm{E}} = H_{\mathrm{m}}^{\mathrm{E}}$, $V_{\mathrm{m}}^{\mathrm{E}}$ are given elsewhere [43]. (ii) It is assumed that only consecutive linear association occurs, which is described by a chemical equilibrium constant ( $K_A$ ) independent of the chain length of the associated species (1-alkanols), according to the equation:

$$A_m + A \leftrightarrow A_{m+1} \quad (9)$$

with *m* ranging from 1 to $\infty$. The cross-association between a self-associated species $A_m$ and a non self-associated compound *B* (benzonitrile in this case) is represented by

$$A_m + B \xleftrightarrow{K_{AB}} A_m B \quad (10)$$

The association constants ( $K_{AB}$ ) of equation (10) are also assumed to be independent of the chain length. Equations (9) and (10) are characterized by $\Delta h_i^*$, the enthalpy of the reaction that corresponds to the hydrogen-bonding energy, and by the volume change ( $\Delta v_i^*$ ) related to the formation of the linear chains. (iii) The $X_{m,phys}^E$ term is derived from the Flory's equation of state [44], which is assumed to be valid not only for pure compounds but also for the mixture [45,46].

$$\frac{\bar{P}_i \bar{V}_i}{\bar{T}_i} = \frac{\bar{V}_i^{1/3}}{\bar{V}_i^{1/3} - 1} - \frac{1}{\bar{V}_i \bar{T}_i} \quad (11)$$

where i = A,B or M (mixture). In equation (11), $\bar{V}_i = V_i / V_i^*$; $\bar{P}_i = P / P_i^*$; $\bar{T}_i = T / T_i^*$ are the reduced volume, pressure and temperature respectively. The pure component reduction parameters $V_i^*$, $P_i^*$, $T_i^*$ are determined from *P-V-T* data (density, $\alpha_p$, isobaric thermal expansion coefficient, and isothermal compressibility, $\kappa_T$ ), and association parameters [45,46]. The reduction parameters for the mixture $P_M^*$ and $T_M^*$ are calculated from mixing rules [45,46]. The total relative molecular volumes and surfaces of the compounds were calculated additively on basis the group volumes and surfaces recommended by Bondi [35].

## 5. Adjustment of model parameters

### *5.1 DISQUAC interaction parameters*

In the framework of DISQUAC, the systems under study are regarded as possessing the following types of surfaces: (i) type a, aliphatic ($CH_3$, $CH_2$, in *n*-alkanes, or aromatic hydrocarbon, or aromatic nitrile, or 1-alkanols); (ii) type n (CN in aromatic nitrile); (iii) type s (s = b, $C_6H_6$, or $C_6H_5$ in aromatic hydrocarbons or nitriles; s = c-$CH_2$ in cycloalkanes; s = h, OH in 1-alkanols).

The general procedure applied in the estimation of the interaction parameters have been explained in detail in earlier works [40,47]. Final values of our fitted parameters are listed in Table 4. For the sake of clarity, we provide now some remarks.

*5.1.1 Benzonitrile + benzene*

This system is only characterized by the (b,n) contact, which is assumed to be represented by DIS and QUAC interaction parameters.

*5.1.2 Benzonitrile + alkane*

These mixtures are built by three contacts: (b,s), (b,n) and (s,n) (s = a,c). The interaction parameters for the (b,s) contacts are dispersive and have been determined from the investigation of alkyl-benzene + alkane systems [37]. The interaction parameters of the (b,n) contacts are already known and thus only those corresponding to the (s,n) contacts must be determined. As in other many applications, we have used $C_{\mathrm{an},l}^{\mathrm{QUAC}} = C_{\mathrm{cn},l}^{\mathrm{QUAC}}$ ($l$ = 1,2,3). That is, the QUAC coefficients for the (a,n) and (c,n) contacts are independent of the alkane [5,47]. Due to the lack of experimental data, we have also assumed $C_{\mathrm{an},2}^{\mathrm{DIS}} = C_{\mathrm{cn},2}^{\mathrm{DIS}}$.

*5.1.3 Benzonitrile + toluene, or + ethylbenzene*

We have here three contacts: (a,b), (a,n) and (b,n). As the interaction parameters for the (a,b) and (b,n) contacts are known, we have determined the $C_{\mathrm{an},l}^{\mathrm{DIS}}$ (l =1,2) coefficients assuming that the $C_{\mathrm{an},l}^{\mathrm{QUAC}}$ (l =1,2,3) coefficients are equal to those of the benzonitrile + *n*-alkane mixtures.

*5.1.4 Phenylacetonitrile or phenylpropionitrile + alkane*

In systems with *n*-alkanes, we have three contacts: (a,b); (a,n) and (b,n). In mixtures with cyclooctane, the contacts are: (a,b); (a,c); (a,n); (b,c); (b,n) and (c,n). We have merely fitted the $C_{\mathrm{sn},1}^{\mathrm{DIS}}$ coefficients against the LLE data assuming that the remainder parameters for the (a,n) and (b,n) contacts are equal to those of benzonitrile + systems.

*5.1.5 1-Alkanols + benzonitrile*

These mixtures are characterized by six contacts: (a,b), (a,h), (a,n), (b,h), (b,n) and (h,n). The contacts (a,h) and (b,h) are represented by DIS and QUAC interaction parameters, which have been determined previously from the study of 1-alkanol + *n*-alkane [38,48,49], or + toluene mixtures [49,50]. Due to the interaction parameters for the (a,b), (a,n) and (b,n) are also known, only those for the (h,n) contacts must be obtained.

*5.2 Adjustment of ERAS parameters*

Values of $V_{\mathrm{i}}, V_{\mathrm{i}}^{*}$ and $P_{\mathrm{i}}^{*}$ of pure compounds at $T$ = 298.15 K, needed for calculations, have been taken from the literature [22] For the 1-alkanols, $K_{\mathrm{A}}$, $\Delta h_{\mathrm{A}}^{*}$ (= − 25.1 kJ·mol$^{-1}$) and $\Delta v_A^*$ (= – 5.6 cm$^3$·mol$^{-1}$) are known from $H_m^E$ and $V_m^E$ data for the corresponding mixtures with alkanes [51]. These values have been used in many other applications [51]. The binary parameters to be fitted against $H_m^E$ and $V_m^E$ data available in the literature for 1-alkanol + benzonitrile systems are then $K_{AB}$, $\Delta h_{AB}^*$, $\Delta v_{AB}^*$ and $X_{AB}$. They are collected in Table 5.

## 6. Theoretical results

Results from the DISQUAC model on phase equilibria, $H_m^E$ and $C_{pm}^E$ are shown in Tables 6-9 and in Figures 1-7 (see also Table S2 and Figures S1, S2 of supplementary material). Tables 6, 7 and 9 contain relative deviations for pressure, temperature and $H_m^E$, respectively, defined as

$$\sigma_r(F) = \{\frac{1}{N}\sum\left[\frac{F_{exp} - F_{calc}}{F_{exp}}\right]^2\}^{1/2} \quad (F = P,T) \tag{12}$$

$$dev(H_m^E) = \{\frac{1}{N}\sum\left[\frac{H_{m,exp}^E - H_{m,calc}^E}{H_{m,exp}^E(x_1 = 0.5)}\right]^2\}^{1/2} \tag{13}$$

ERAS results on $H_m^E$ and $V_m^E$ for 1-alkanol + benzonitrile systems are shown in Tables 9 and 10 (Figures 6 and 8). Results on SLE from the Ideal Solubility Model are collected in Table 7 (Figures 7 and S2)

## 7. Discussion

Below, we are referring to values of the excess molar functions at 298.15 K and equimolar composition. We also refer to aromatic polar compounds of general formula $C_6H_5-(CH_2)_{n-1}X$ (X = Cl, CN, $NH_2$, OH ) or $C_6H_5-(CH_2)_{n-1}XCH_3$ (X = CO, CHO, COO, O).

We start remarking that the UCST values, ranged between ≈ (280,360) K for systems formed by one aromatic nitrile and one alkane reveal the existence of strong dipolar interactions between nitrile molecules. Accordingly, the $H_m^E$ and $TS_m^E (= H_m^E - G_m^E)$ values for the benzonitrile + cyclohexane system are rather large: 1390 [52] and 380 J·mol$^{-1}$, respectively (the latter value determined using the DISQUAC value for $G_m^E$, 1011 J·mol$^{-1}$). The existence of dipolar interactions is also supported by low positive $C_{pm}^E$ values. Thus, for benzonitrile systems, $C_{pm}^E$ /J·mol$^{-1}$ ·K$^{-1}$ = 4.7 (octane); 5.2 (nonane) [53] or 1.6 (cyclohexane) [54]. As usual, $C_{pm}^{E}$ values increase when the temperature is approaching to the UCST [53,54], as close to the critical point, non-random effects become more important. On the other hand, the very different behaviour of $V_{m}^{E}$ observed for benzonitrile + alkane mixtures must be mentioned. For example, systems with octane or nonane show S-shaped $V_{m}^{E}$ curves, with very low positive values at low nitrile concentrations, say up to $x_1 \approx 0.2$ for the nonane solution, and, at $x_1 = 0.5$, $V_{m}^{E}$ /cm$^3$·mol$^{-1}$ = $-0.298$ (octane), $-0.159$ (nonane) [53]. This means that structural effects are dominant in a large concentration range (see below). For the cyclohexane, mixture, $V_{m}^{E}$ is positive at any composition ($V_{m}^{E}$ = 0.26 cm$^3$·mol$^{-1}$ [54]) which clearly reveals that the dominant effects on $V_{m}^{E}$ arise from interactional effects.

Interestingly, interactions between polar like molecules are stronger when the polar group is attached to an aromatic ring than when the mentioned group is placed in a linear chain. The following $H_{m}^{E}$ values demonstrate this statement; $H_{m}^{E}$(heptane)/J·mol$^{-1}$ = 1492 (acetophenone) [55]; 886 (2-heptanone) [56]; 1361 (benzaldehyde) [57]; 1066 (1-pentanal) [58]; 1154 (ethyl benzoate [59], 528 (ethyl hexanoate) [60]; 1278 (methoxybenzene [61]; 260 (1-methoxypentane) [62]; 697 (chlorobenzene) [63]; 396 (1-chlorohexane) [64]; 962 (1-hexylamine) [65]; 750 (1-hexanol + heptane, $T$ = 318.15 K) [66]; or $H_{m}^{E}$ (cyclohexane)/J·mol$^{-1}$ = 1101 (pentanitrile) [67]; 1390 (benzonitrile) [52]. Aniline or phenol + heptane mixtures show miscibility gaps with rather high UCST values: 343.1 K [68] and 327.3 K [69], respectively. This behaviour can be explained in terms of an enhancement of dipolar interactions due to the presence of the $C_6H_5$- and X groups within the same molecule. These *intramolecular effects* are the so-called proximity effects. In fact, if the aromatic ring and the polar group belong to different molecules (as in linear polar compound + $C_6H_6$ mixtures), *intermolecular effects* come into play, particularly interactions between unlike molecules and $H_{m}^{E}$(aromatic polar compound + alkane) > $H_{m}^{E}$ (linear polar compound + $C_6H_6$). For comparison, we provide some experimental results (see above); $H_{m}^{E}$($C_6H_6$)/J·mol$^{-1}$ = 138 (2-propanone) [70]; $-171$ (2-hexanone) [71]; 54 (propanal); $-82$ (pentanal) [72]; 84 (ethyl ethanoate) [59]; 627 (1-hexylamine, $T$ = 303.2 K) [73]; $-112$ (pentanenitrile) [74]; 79 (1-chlorohexane) [75].

Intramolecular effects between two equal or different groups in the same molecule also leads to stronger interactions in mixtures of alkane with linear polyoxaalkane [76], $\alpha,\beta$-dichloroalkane [77]; morpholine [78]; amine-ketone [79], or alkoxyethanol [5].

*7.1 Dependence of proximity effects on the separation between the polar and phenyl groups*

We note that for X = CN, UCST($n$-$C_8$)/K = 356.3 ($n$ = 3) > 352.6 ($n$ = 2) (Table 3) > 283.2 ($n$ = 1) [80]. That is, interactions between aromatic nitrile molecules become weaker when the separation between the $C_6H_5-$ and CN groups decreases. This behaviour is different to that observed for alkanenitrile + heptane systems, as $H_m^E$($n$-$C_7$)/J·mol$^{-1}$ = 1561 (propanenitrile) [81] > 1304 (butanenitrile) > 1129 (pentanenitrile) [67] and UCST(ethanenitrile + $n$-$C_7$) = 358 K [82]. An useful magnitude to roughly estimate relative changes in intermolecular forces of homomorphic compounds is $\Delta\Delta H_{vap}$ , defined as [8,76]:

$$\Delta\Delta H_{vap} = \Delta H_{vap} \text{ (compound with a given polar group, X) } - \Delta H_{vap} \text{(homomorphic hydrocarbon)} \quad (14)$$

In this equation $\Delta H_{vap}$ is the standard enthalpy of vaporization at 298.15 K. Some values for nitriles are now given, $\Delta H_{vap}$/kJ·mol$^{-1}$ = 60.5 (phenylacetonitrile) [83]; 55.5 (benzonitrile) [84]; 36.0 (propanenitrile), 43.6 (pentanenitrile), 56.8 (octanenirile) [85]. Using $\Delta H_{vap}$ values compiled in [85] for the isomeric hydrocarbons, we obtain $\Delta\Delta H_{vap}$/kJ·mol$^{-1}$ = 18.8 (phenylacetonitrile) > 17.5 (benzonitrile) and $\Delta\Delta H_{vap}$/kJ·mol$^{-1}$ = 21.9 (propanenitrile) > 17.0 (pentanenitrile) > 16.4 (hexanenitrile) > 15.3 (octanenitrile). These values are in agreement with the trends stated above: (i) for isomeric molecules, interactions between nitrile molecules are stronger when the polar group is attached to an aromatic ring; (ii) these interactions are stronger in phenylacetonitrile than in benzonitrile, while become weaker when the size of the alkanenitrile is increased. For systems containing other aromatic polar compounds, there is a variety of different behaviours depending on the considered group. Thus, UCST(X = CO; $n$-$C_{10}$)/K = 301.6 ($n$ =2) > 284.5 ($n$ = 3) [9] > 277.4 ($n$ =1) [8]; and $H_m^E$( $n$-$C_7$)/ J·mol$^{-1}$ = 1680 ($n$ =2) [55] > 1604 ($n$ = 3) [61] > 1492 ($n$ = 1) [55]. A similar trend is encountered for aromatic chloroalkanes (X = Cl), $H_m^E$ ($n$-$C_7$)/J·mol$^{-1}$= 1548 ($n$ = 2) > 1429 ($n$ = 3) > 1082 ($n$ = 4) > 697 ($n$ =1) [63]. Systems with X = O or CHO behave differently. $H_m^E$( $n$-$C_7$, X = O)/J·mol$^{-1}$ = 1278 ($n$ =1) > 1213 ($n$ = 2) [61] > 1148 ($n$ =3) [86]; and $H_m^E$(X = CHO; $n$-$C_7$)/J·mol$^{-1}$ = 1480 ($n$ = 2) > 1367 ($n$ =1) > 1061 ($n$ = 3) [57]. In the case of systems with self-associated compounds (X = $NH_2$, OH), interactions between polar molecules are stronger when $n$ = 1, as UCST(X = $NH_2$)/K = 343.1 ($n$-$C_7$, $n$ =1) [68] > 280.1 ($n$-$C_{10}$, $n$ = 2) [12], and UCST(X = OH; $n$-$C_{10}$)/K = 336.5 ($n$ =1) [87] > 335.8 ($n$ =2) [13].

*7.2 The effect of replacing octane by 2,2,4-trimethylpentane in systems with a polar compound*

We note that UCST(phenylacetonitrile)/K = 359.8 ($i$-$C_8$) > 352.5 ($n$-$C_8$) (Table 3). The replacement of $n$-$C_8$ by $i$-$C_8$ also leads to higher values for systems with phenyl acetone (293.4 K ($n$-$C_8$, estimated result); 298.5 K ($i$-$C_8$)) [9]; phenol (329.5 K ($n$-$C_8$) [69]; 339.1 ($i$-$C_8$) [88]); aniline (345.1 K ($n$-$C_8$); 353.1 K ($i$-$C_8$)) [88] or $\varepsilon$-caprolactam (354.5 K ($n$-$C_8$); 362.3 K ($i$-$C_8$)) [89].

The opposite trend is observed for systems involving methanol (339.3 K ($n$-$C_8$) [27]; 315.6 K ($i$-$C_8$) [90]); 2-methoxyethanol (327.9 K ($n$-$C_8$) [32]; 319.2 K ($i$-$C_8$) [4]); 2-(2-ethoxyethoxy)ethanol (294.7 K ($n$-$C_8$) [91]; 290.2 K ($i$-$C_8$) [4]); *N,N*-dimethylacetamide (DMA) (314.0 K ($n$-$C_8$) [92]; 290.6 K ($i$-$C_8$) [93]); or *N*-methylpyrrolidone (NMP) (328.5. K ($n$-$C_8$) [94]; 326.9 K ($i$-$C_8$) [95]).

It seems that the replacement of $n$-$C_8$ by $i$-$C_8$ in systems with a given aromatic polar compound leads to increased values of UCST. For other polar compounds, the result depends on the nature of the compound (see UCST values for NMP and $\varepsilon$-caprolactam) and of the branching of the alkane [93].

*7.3 The effect of replacing a linear alkane by an isomeric cycloalkane in systems with a polar compound*

The substitution of a linear alkane by an isomeric cyclolkane leads to lower UCST values for systems containing *N,N*-dimethylformamide (337.7 K ($n$-$C_6$) [96]; 320.0 K (c-$C_6$) [97]); DMA (314.0 K ($n$-$C_8$) [92]; 299.3 K (c-$C_8$) [93]); NMP (324.6 K ($n$-$C_6$) [94]; 283.1 K (c-$C_6$) [98]; 328.5 K ($n$-$C_8$) [94]; 291.9 K (c-$C_8$) [98]); 2ME ((311.2 K ($n$-$C_6$); 294.6 K (c-$C_6$)) [32]; acetic anhydride ((335.0 K ($n$-$C_6$); 323.5 K (c-$C_6$)) [99]; 2-phenoxyethanol ((365 K (estimated value, $n$-$C_6$); 314.9 K (c-$C_6$)) [14], or phenylacetonitrile (350.1 K ($n$-$C_8$), 310.4 K (c-$C_8$) Table 3). One can conclude that UCST decreases independently of the polar substance when a linear alkane is replaced by an isomeric cycloalkane.

*7.4 Aromatic nitrile + aromatic hydrocarbon*

We have already mentioned that M-shaped $H_m^E$ curves have been encountered for benzonitrile + $C_6H_6$, or + $C_7H_8$ systems [23,24], and that such anomalous concentration dependence of $H_m^E$ has been ascribed to charge-transfer complex formation between benzonitrile and the aromatic hydrocarbon [23]. In order to elucidate this point, we have calculated the excess molar internal energies at constant volume, $U_{Vm}^E$, from the equation [29,100]:

$$U_{\mathrm{Vm}}^{\mathrm{E}} = H_{\mathrm{m}}^{\mathrm{E}} - \frac{\alpha_{\mathrm{p}}}{\kappa_{\mathrm{T}}} T V_{\mathrm{m}}^{\mathrm{E}} \tag{15}$$

where $\frac{\alpha_p}{\kappa_T} T V_m^{\mathrm{E}}$ is the so-called equation of state (eos) contribution to $H_{\mathrm{m}}^{\mathrm{E}}$, $\alpha_{\mathrm{p}}$ and $\kappa_T$ are, respectively, the isobaric thermal expansion coefficient and the isothermal compressibility of the mixture. Values of $\alpha_p$ and $\kappa_T$ have been determined assuming ideal behaviour for the considered systems with regards to these properties ($F = \Phi_1 F_1 + \Phi_2 F_2$; $F_i$ is the property of the pure compound *i*). Results are plotted in Figure 9. We note that the $U_{\mathrm{Vm}}^{\mathrm{E}}$ curves show a normal concentration dependence. This suggests that the anomalous behavior of $H_{\mathrm{m}}^{\mathrm{E}}$, if exists, may be ascribed to structural effects. The existence of such effects is supported by: (i) the different signs of the $H_{\mathrm{m}}^{\mathrm{E}}$ and $V_{\mathrm{m}}^{\mathrm{E}}$ functions [101,102]; (ii) the decrease of $V_{\mathrm{m}}^{\mathrm{E}}$ when the temperature is increased [23,103]. For the benzene mixture, $H_{\mathrm{m}}^{\mathrm{E}}$ = 32 J·mol$^{-1}$; $V_{\mathrm{m}}^{\mathrm{E}}$ = − 0.116 cm$^3$·mol$^{-1}$; $dV_{\mathrm{m}}^{\mathrm{E}} / dT$ = − 4.9 10$^{-3}$ cm$^3$·mol$^{-1}$·K$^{-1}$ [23]; and for the ethylbenzene solution, $H_{\mathrm{m}}^{\mathrm{E}}$ = 105 J·mol$^{-1}$ [25]; $V_{\mathrm{m}}^{\mathrm{E}}$ = − 0.369 cm$^3$·mol$^{-1}$; $dV_{\mathrm{m}}^{\mathrm{E}} / dT$ = − 1.08 10$^{-3}$ cm$^3$·mol$^{-1}$·K$^{-1}$ [103]. For the toluene mixture, $H_{\mathrm{m}}^{\mathrm{E}}$ = − 15 J·mol$^{-1}$; $V_{\mathrm{m}}^{\mathrm{E}}$ = − 0.294 cm$^3$·mol$^{-1}$; $dV_{\mathrm{m}}^{\mathrm{E}} / dT$ = − 2.96 10$^{-3}$ cm$^3$·mol$^{-1}$·K$^{-1}$ [23]. The $V_{\mathrm{m}}^{\mathrm{E}}$ variation indicates that structural effects decrease in the order: ethylbenzene > toluene > benzene. Interestingly, the $H_{\mathrm{m}}^{\mathrm{E}}$ variation is not consistent with the fact that the positive contribution to $H_{\mathrm{m}}^{\mathrm{E}}$ from the disruption of dipolar interactions between benzonitrile molecules should increase with the alkylation of the aromatic hydrocarbon. However, $U_{\mathrm{Vm}}^{\mathrm{E}}$/J·mol$^{-1}$ changes as expected (Figure 9): 78 (benzene) < 100 (toluene) < 141 (ethylbenzene). Finally, it must be remarked that the SLE phase diagrams of the benzonitrile + benzene, or + toluene systems show merely a simple eutectic point [104] (Figure S2), what also supports that no complex formation between compounds takes place. In fact, the $H_{\mathrm{m}}^{\mathrm{E}}$ curve of the $CCl_4$ + benzene system show the typical parabolic shape [105], but the SLE phase diagram reveals the existence of a complexes of the 1:1 or 1:2 type [106,107].

On the other hand, $C_{\mathrm{pm}}^{\mathrm{E}}$ (benzonitrile)/J·mol$^{-1}$·K$^{-1}$ changes as follows: − 0.59 (benzene) < 1.05 (toluene) [23] < 2.12 (ethylbenzene) [103]. Negative $C_{\mathrm{pm}}^{\mathrm{E}}$ values are encountered in systems where dispersive interactions are dominant (− 3.34 J·mol$^{-1}$·K$^{-1}$ for the benzene + heptane mixture [108]). Positive $C_{\mathrm{pm}}^{\mathrm{E}}$ values reveal that dipolar interactions become progressively more relevant with the increasing of the aliphatic surface of the aromatic hydrocarbon. This picture is still valid when excess molar volumes at constant volume, $C_{\mathrm{Vm}}^{\mathrm{E}}$, are considered as the values of this magnitude, determined from the equation:

$$C_{\mathrm{Vm}}^{\mathrm{E}} = C_{\mathrm{pm}}^{\mathrm{E}} - \frac{\alpha_{\mathrm{p}}^{2}}{\kappa_{\mathrm{T}}} T V_{\mathrm{m}}^{\mathrm{E}} \quad (16)$$

are close to those of $C_{\mathrm{pm}}^{\mathrm{E}}$: $-0.54$ (benzene); 1.16 (toluene), 2.25 (ethylbenzene) J·mol$^{-1}$·K$^{-1}$. In conclusion, all these features suggest that interactions in the present systems are mainly of dispersive type.

For a given aromatic hydrocarbon, say toluene, $V_{\mathrm{m}}^{\mathrm{E}}$/cm$^3$·mol$^{-1}$ values change in the order: $-0.389$ phenylacetonitrile [109] < $-0.298$ (benzonitrile) [23], while dipolar interactions are stronger in phenylacetonitrile mixtures. This suggests that structural effects are more relevant in phenylacetonitrile systems.

*7.5 1-Alkanol + benzonitrile*

The main features of 1-alkanol + nitrile mixtures have been examined in detail previously and will not be repeated here [20]. We merely remark some important points regarding 1-alkanol + benzonitrile systems. (i) The rather large and positive $H_{\mathrm{m}}^{\mathrm{E}}$ values together with the nearly symmetrical $H_{\mathrm{m}}^{\mathrm{E}}$ curves of these systems (Figure 6) point out to the existence of dipolar interactions. (ii) $H_{\mathrm{m}}^{\mathrm{E}}$ and $V_{\mathrm{m}}^{\mathrm{E}}$ are of opposite sign (Tables 9,10), 976 J·mol$^{-1}$ and $-0.358$ cm$^3$·mol$^{-1}$, respectively for the methanol system [22], which reveals the existence of strong structural effects in these mixtures. (iii) In addition, both $H_{\mathrm{m}}^{\mathrm{E}}$ and $V_{\mathrm{m}}^{\mathrm{E}}$ increase with the 1-alkanol size. This is due to interactions between unlike molecules become weaker when the alkanol size increases, and to the positive contribution from the breaking of interactions between benzonitrile molecules increases at the same condition [20]. (iv) Interestingly, for a given 1-alkanol, we note that $H_{\mathrm{m}}^{\mathrm{E}}$(toluene) < $H_{\mathrm{m}}^{\mathrm{E}}$(benzonitrile). For example, $H_{\mathrm{m}}^{\mathrm{E}}$(methanol)/J·mol$^{-1}$ = 622 (toluene) [110] < 976 (benzonitrile) [22], or $H_{\mathrm{m}}^{\mathrm{E}}$(1-propanol)J·mol$^{-1}$ = 880 (toluene) [110] < 1454 (benzonitrile) [22]. This means that benzonitrile is a more active molecule than toluene when breaking the alkanol network, but also 1-alkanols are good breakers of the dipolar interactions between benzonitrile molecules.

We have conducted some calculations in order to determine the Kirkwood's correlation factor, $g_{\mathrm{K}}$, from data available in the literature on permittivity, density and refractive indices for methanol or 1-propanol + benzonitrile mixtures [111,112]. The equation used for $g_{\mathrm{K}}$ is [113-115]:

$$g_{\mathrm{K}} = \frac{9k_{\mathrm{B}}TV_{\mathrm{m}}\varepsilon_0(\varepsilon_{\mathrm{r}} - \varepsilon_{\mathrm{r}}^{\infty})(2\varepsilon_{\mathrm{r}} + \varepsilon_{\mathrm{r}}^{\infty})}{N_{\mathrm{A}}\mu^2\varepsilon_{\mathrm{r}}(\varepsilon_{\mathrm{r}}^{\infty} + 2)^2} \tag{17}$$

where the symbols have the usual meaning [116]. Results shown in Figure 10 reveal that the system structure changes similarly for both solutions. Nevertheless, cooperative effects which lead to an increasing of the effective dipole moment of the mixture are weaker in the 1-propanol system. Accordingly, the excess permittivity is positive for the methanol solution (0.4) and negative ($-1$) for the mixture with 1-propanol. On the other hand, comparison of $g_{\mathrm{K}}$ for 1-propanol + benzonitrile, or + toluene [117,118] systems (Figure 10) indicates that the effects related to the alcohol self-association are much relevant in the toluene mixture as $g_{\mathrm{K}}$ increases very sharply with the alkanol concentration.

*7.6 DISQUAC results*

Regarding phase equilibria, VLE, LLE or SLE, the model provides rather good results (Tables 6 and 7; Figures 1, 2, 7, S1 and S2). As usually, the coordinates of the critical and eutectic points are represented in the correct range of temperature and composition (Tables 8 and S2). We note that the theoretical $H_{\mathrm{m}}^{\mathrm{E}}$ values obtained using DISQUAC for benzonitrile + benzene, or + toluene systems are somewhat poor (Table 9). This can be ascribed to the low $|H_{\mathrm{m}}^{\mathrm{E}}|$values of the mentioned mixtures in conjunction with the unusual shape of the corresponding curves reported by several authors. A matter already discussed in detail. In spite of this, we must remark the excellent $C_{\mathrm{pm}}^{\mathrm{E}}$ results given by the model for benzonitrile + hydrocarbon systems (Figures 4 and 5). On the other hand, DISQUAC provides rather good $H_{\mathrm{m}}^{\mathrm{E}}$ results for systems formed by benzonitrile and ethylbenzene, cyclohexane, methanol or 1-propanol (Table 9, Figures 3 and 6). The larger discrepancies between experimental values and theoretical calculations encountered for the ethanol solution (Table 9) are discussed below.

*7.7 ERAS results*

The $H_{\mathrm{m}}^{\mathrm{E}}$ function of 1-alkanol + benzonitrile mixtures is well described by the model (Table 9, Figure 6), which means that association/solvation effects are relevant in these systems. Calculations are conducted using $\Delta h_{\mathrm{AB}}^{*} = -12$ kJ·mol$^{-1}$ (Table 5) a very different value to those obtained previously for the enthalpy of the 1-alkanol-benzonitrile interactions [20]: $-27.1$ (methanol), $-25.4$ (ethanol); $-24.6$ (1-propanol) kJ·mol$^{-1}$. The poorer results obtained for the mixture with 1-propanol may be related to a decrease of association/solvation effects in this system compared to those in the methanol solution. This is in agreement with our calculations on $g_K$. Regarding $V_m^E$, its variation with the alkanol size is well reproduced by the model (Table 10). However, results become poorer for systems with 1-propanol or 1-pentanol (Figure 10), indicating that structural effects are not correctly represented for such solutions. We must remark that our results are obtained using a set of parameters which smoothly change with the molecular structure. Parameters previously reported for these systems change more erratically with the chain length of the 1-alkanol [22]. Thus, $K_{AB}$ = 19 (methanol); 6 (ethanol); 18 (1-propanol); $\Delta h_{AB}^*$/kJ·mol$^{-1}$ = $-9.5$ (methanol); $-3.8$ (ethanol); $-10.5$ (1-propanol); $\Delta v_{AB}^*$/cm$^3$·mol$^{-1}$ = $-10$ (methanol); $-15$ (ethanol); $-11$ (1-propanol) and $X_{AB}$/J·cm$^{-3}$ = 19 (methanol); 4 (ethanol); 8 (1-propanol).

*7.8 DISQUAC interaction parameters*

The (b,n) contact in the benzonitrile + benzene mixture is described by DIS and QUAC interchange coefficients, although interactions in this systems are mainly of dispersive type (see above). However, calculations show that the temperature dependence of the thermodynamic properties of benzonitrile + alkane mixtures is better represented using also $C_{bn,l}^{QUAC}$ ($l$ =1,2,3) coefficients. Nevertheless, we remark the low values of the QUAC parameters for $l$ =1,2.

A comment with regard to the dependence of the $C_{an,1}^{DIS}$ coefficient with *n*-alkane size is needed. We have encountered the same behaviour in other many systems previously investigated as those containing one *n*-alkane and *N,N*-dialkylamide [47], or pyridine [119], or benzylalcohol [13], or acetophenone [8]. Calculations show that the first Gibbs dispersive parameter must be assumed to be dependent on the alkane size (Table 4) in order to provide correct values of ($x_{1c}$, $T_c$) (Table 8). This can be explained taking into account that DISQUAC is a mean field theory and that, therefore, theoretical calculations on LLE are developed under the assumption that $G_m^E$ is an analytical function close to the critical point. However, this is a wrong assumption as it is known that, at temperatures close to $T_c$, thermodynamic functions are expressed in terms of scaling laws with universal critical exponents and universal scaling functions [29]. As a consequence, theoretical LLE curves are more rounded than the experimental ones at temperatures not far from the UCST (Figures 1,2). Moreover, the calculated critical temperatures are higher than the experimental values at UCST and lower than the experimental results at the LCST [29] (lower critical solution temperature). In spite of these shortcomings of mean field theories, it is remarkable that DISQUAC correctly describes the change in the symmetry of the LLE curves for benzonitrile mixtures when the alkane size is increased (Figure 2).

Finally, the poorer results provided by the model for the ethanol solution (Table 9) can be ascribed to the similar $\alpha_{ai}$ values for ethanol (0.2961) and benzonitrile (0.2912). In fact, the $H_m^{E,DIS}$ contribution is here the result of the sum of six terms (there are 6 contacts in the ethanol mixture), three of them are dependent on the very small ($\alpha_{a1} - \alpha_{a2}$) difference. This makes very difficult to obtain a set of interaction parameters smoothly dependent on the molecular structure. Better results are obtained using $C_{hn,2}^{DIS} = -53$ and $C_{hn,2}^{QUAC} = 24$ ($dev(H_m^E) = 0.016$), but these interchange coefficients largely differ from those listed in Table 4 for methanol or 1-propanol mixtures.

## 8. Conclusions

LLE data for the systems phenylacetonitrile + *n*-$C_7$, + *n*-$C_8$,+ *n*-$C_9$, + *c*-$C_8$, or + *i*-$C_8$ and for 3-phenylpropionitrile + *n*-$C_7$, or + *n*-$C_8$, have been reported. Dipolar interactions between aromatic nitrile molecules become weaker in the sequence: 3-phenylpropionitrile > phenylacetonitrile > benzonitrile. Aromatic nitrile + alkane, + aromatic hydrocarbon, or + 1-alkanol mixtures have been studied using DISQUAC. The interchange coefficients for contacts involving the aromatic nitrile group have been reported. DISQUAC describes correctly any type of phase equilibria, $C_{pm}^E$ of benzonitrile + hydrocarbon mixtures and $H_m^E$ of benzonitrile + cyclohexane, or + 1-alkanol. Differences between theoretical $H_m^E$ values and experimental data for other systems have been rationalized. The ERAS model has also been applied to 1-alkanol + benzonitrile mixtures. ERAS represents well $H_m^E$ of these solutions and more poorly the $V_m^E$ curves of systems with longer 1-alkanols, which has been ascribed to the existence of structural effects.

## Acknowledgements

The authors gratefully acknowledge the financial support received from the Consejería de Educación y Cultura of Junta de Castilla y León, under Project BU034U16. F. Hevia gratefully acknowledges the grant received from the program ‘Ayudas para la Formación de Profesorado Universitario (convocatoria 2014), de los subprogramas de Formación y de Movilidad incluidos en el Programa Estatal de Promoción del Talento y su Empleabilidad, en el marco del Plan Estatal de Investigación Científica y Técnica y de Innovación 2013-2016, de la Secretaría de Estado de Educación, Formación Profesional y Universidades, Ministerio de Educación, Cultura y Deporte, Gobierno de España’.

## 9. References

[1] A.L. McClellan, Tables of Experimental Dipole Moments, Vols., 1,2,3, Rahara Enterprises, El Cerrito, US, 1974.

[2] H.-S. Byun, J. Supercrit. Fluids 120 (2017) 218-225.

[3] F. Aguilar, F.E.M. Alaoui, J.J. Segovia, M.A. Villamañán, E.A. Montero, J. Chem. Thermodyn. 42 (2010) 28-37.

[4] F. J. Carmona, J.A. González, I. García de la Fuente, J.C. Cobos, J. Chem. Eng. Data 44 (1999) 892-895.

[5] J.A. González, J.C. Cobos, I. García de la Fuente, V.R. Bhethanabotla, S.W. Campbell. Phys. Chem. Chem. Phys. 3 (2001) 2856-2865.

[6] S. Delcros, J.R. Quint, J.-P.E. Grolier, H.V. Kehiaian, Fluid Phase Equilib. 113 (1995) 1-19.

[7] J. Gmehling, J. Li, M. Schiller, Ind. Eng. Chem. Res. 32 (1993) 178-193.

[8] J.A. González, C. Alonso-Tristán, I. García de la Fuente, J.C. Cobos, Fluid Phase Equilib. 391 (2015) 39-48.

[9] J.A. González, C. Alonso-Tristán, I. García de la Fuente, J.C. Cobos, Fluid Phase Equilib. 421 (2016) 49-58.

[10] C. Alonso-Tristán, J.A. González, F. Hevia, I. García de la Fuente, J.C. Cobos, J. Chem. Eng. Data 62 (2017) 988-944.

[11] J.A. González, C. Alonso-Tristán, I. García de la Fuente, J.C. Cobos. Fluid Phase Equilib. 366 (2014) 61-68.

[12] C. Alonso-Tristán, J.A. González, I. García de la Fuente, J.C. Cobos, J. Chem. Eng. Data 59 (2014) 2101-2105.

[13] J.A. González, C. Alonso-Tristán, I. García de la Fuente, J. C. Cobos, J. Chem. Eng. Data, 57 (2012) 1186-1191.

[14] V. Alonso, M. García, J. A. González, I. García de la Fuente, J.C. Cobos, Thermochim. Acta 521 (2011) 107-111.

[15] J.A. González, I. Mozo, I. García de la Fuente, J.C. Cobos, Can. J. Chem. 83 (2005) 1812-1825.

[16] J.A. González, I. García de la Fuente, J.C. Cobos, Ber. Bunsenges. Phys. Chem. 100 (1996) 1746-1751.

[17] H.V. Kehiaian, Fluid Phase Equilib. 13 (1983) 243-252.

[18] B. Marongiu, B. Pittau, S. Porcedda, Thermochim. Acta 221 (1993) 143-162.

[19] J.A. González, F. Hevia, A. Cobos, I. García de la Fuente, C. Alonso-Tristán, Thermochim. Acta 605 (2015) 121-129.

[20] J.A. González, I. García de la Fuente, J.C. Cobos, C. Alonso-Tristán, L.F. Sanz. Ind. Eng. Chem. Res. 54 (2015) 550-559.

[21] A. Heintz, Ber. Bunsenges. Phys. Chem. 89 (1985) 172-181.
[22] T.M. Letcher, P.K. Naicker, J. Chem. Thermodyn. 33 (2001) 1035-1047.
[23] E. Wilhelm, W. Egger, M. Vencour, A.H. Roux, M. Polednicek, J.-P.E. Grolier, J. Chem. Thermodyn. 30 (1998) 1509-1532.
[24] S. Horstmann, H. Gradeler, R. Bölts, J. Gmehling, J. Chem. Eng. Data 44 (1999) 539-543.
[25] T.M. Letcher, P.K. Naicker, J. Chem. Thermodyn. 33 (2001) 1027-1033.
[26] CIAAW, Atomic weights of the elements 2015, ciaaw.org/atomic-weights.htm, 2015.
[27] J.B. Ott, I.F. Hölscher, G.M. Schneider, J. Chem. Thermodyn. 18 (1986) 815-826.
[28] U. Haarhaus, G.M. Schneider, J. Chem. Thermodyn. 20 (1988) 1121-1129.
[29] J.S. Rowlinson, F.L. Swinton, Liquids and Liquid Mixtures, 3rd ed., Butterworths, London, (1982).
[30] H.E. Stanley, Introduction to Phase Transitions and Critical Phenomena, Clarendon Press, Oxford, (1971).
[31] J.P. Novak, J. Matous, J. Pick, Liquid-Liquid Equilibria, Elsevier: Amsterdam, (1987).
[32] M.A. Rubio, J.A. González, I. García de la Fuente, J.C. Cobos, Fluid Phase Equilib. 143 (1998) 111-123.
[33] P.R. Bevington, Data Reductions and Error Analysis for the Physical Sciences. McGraw-Hill Book Co., London, (1969).
[34] E.A. Guggenheim, Mixtures, Oxford University Press, Oxford, (1952).
[35] A. Bondi, Physical Properties of Molecular Crystals, Liquids and Glasses, Wiley, New York, (1968).
[36] H.V. Kehiaian, J.-P.E. Grolier, G.C. Benson, J. Chim. Phys. 75 (1978) 1031-1048.
[37] A. Cannas, B. Marongiu, S. Porcedda, Thermochim. Acta 311 (1998) 1-19.
[38] J.A. González, I. García de la Fuente, J.C. Cobos, C. Casanova, Ber. Bunsenges. Phys. Chem. 95 (1991) 1658-1668.
[39] J.A. González, I. García de la Fuente, J.C. Cobos, C. Casanova, A. Ait-Kaci, Fluid Phase Equilib. 112 (1995) 63-87.
[40] J.A. González, I. García de la Fuente, J.C. Cobos, Fluid Phase Equilib. 168 (2000) 31-58.
[41] I. Prigogine, R. Defay, Chemical Thermodynamics, Longmans, Green and Co., Norwich, (1954).
[42] J.M. Prausnitz, Molecular Thermodynamics of Fluid Phase Equilibria, Prentice-Hall, Englewood Cliffs, NJ. (1969)
[43] J.A. González, S. Villa, N. Riesco, I. García de la Fuente, J.C. Cobos, Can. J. Chem. 81 (2003) 319-329.
[44] P.J. Flory, J. Am. Chem. Soc., 87 (1965) 1833-1838.

[45] A. Heintz, P.K. Naicker, S.P. Verevkin, R. Pfestorf, Ber. Bunsenges. Phys. Chem. 102 (1998) 953-959.

[46] A. Heintz, D. Papaioannou, Thermochim. Acta, 310 (1998) 69-76.

[47] J.A. González, J.C. Cobos, I. García de la Fuente, Fluid Phase Equilib. 224 (2004) 169-183.

[48] J.A. González, I. García de la Fuente, J.C. Cobos, C. Casanova, U. Domanska. Ber. Bunsenges. Phys. Chem. 98 (1994) 955-959.

[49] U. Domanska, J.A. González, Fluid Phase Equilib. 119 (1996) 131-151.

[50] J.A. González, I. García de la Fuente, J.C. Cobos, C. Casanova, Fluid Phase Equilib. 93 (1994) 1-22.

[51] J.A. González, I. Mozo, I. García de la Fuente, J.C. Cobos, N. Riesco, J. Chem. Thermodyn. 40 (2008) 1495-1508.

[52] R. Tanaka, S. Murakami, R. Fujishiro, J. Chem. Thermodyn. 6 (1974) 209-218.

[53] Y. Lei, Z. Chen, X. An, M. Huang, W. Shen, J. Chem. Eng. Data 55 (2010) 4154-4161.

[54] R. Tanaka, T. Nakamichi, S. Murakami, J. Solution Chem. 14 (1985) 795-803.

[55] O. Urdaneta, S. Haman, Y.P. Handa, G.C. Benson, J. Chem. Thermodyn. 11 (1979) 851-856. (d1)

[56] O. Urdaneta, Y.P. Handa, G.C. Benson, J. Chem. Thermodyn. 11 (1979) 857-860.

[57] I. Ferino, B. Marongiu, V. Solinas, S. Torrazza, Thermochim. Acta 57 (1982) 147-154.

[58] I. Ferino, B. Marongiu, V. Solinas, S. Torrazza, H.V. Kehiaian, Fluid Phase Equilib. 12 (1983) 125-142.

[59] J.-P.E. Grolier, D. Ballet, A. Viallard, J. Chem. Thermodyn. 6 (1974) 895-908.

[60] M. Vidal, J. Ortega, J. Plácido. J. Chem. Thermodyn. 29 (1997) 47-74.

[61] B. Marongiu, A. Piras, S. Porcedda, E. Tuveri, J. Thermal Anal. Calorim. 92 (2008) 137-144.

[62] K. Sosnkowska-Kehiaian, R. Hryniewicz, H.V. Kehiaian, Bull. Acad., Polon. Sci., Ser. Sci. Chem. 17 (1969) 185-190.

[63] B. Marongiu, A. Piras, S. Porcedda, E. Tuveri, J. Therm. Anal. Calorim. 91 (2008) 37-46

[64] M.I. Paz-Andrade, R. Bravo, M. García, J.-P.E. Grolier, J. Chim. Phys. Phys.-Chim. Biol. 76 (1979) 51-56.

[65] E. Matteoli, L. Lepori, A. Spanedda, Fluid Phase Equilib. 212 (2003) 41-52.

[66] H. Kirss, M. Kuus, E. Siimer, L. Kudryavtseva, Thermochim. Acta 265 (1995) 45-54.

[67] B. Marongiu, S. Porcedda, J. Chem. Eng. Data 35 (1990) 172-174.

[68] H. Matsuda, K. Ochi, K. Kojima, J. Chem. Eng. Data 48 (2003) 184-189.

[69] H. Matsuda, M. Fujita, K. Ochi, J. Chem. Eng. Data 48 (2003) 1076-1080.

[70] Y. Akamatsu, H. Ogawa, S. Murakami, Thermochim. Acta 113 (1987) 141-150.

[71] B. Marongiu, S. Dernini, L. Lepori, E. Matteoli, Thermochim. Acta 149 (1989) 111-121.
[72] B. Marongiu, Thermochim. Acta, 95 (1985) 105-110.
[73] I. Velasco, S. Otín, C. Gutiérrez Losa, J. Chim. Phys. Phys.-Chim. Biol. 75 (1978) 706-708.
[74] B. Marongiu, S. Porcedda, M. Marrocu, D. Falconieri, A. Piras, J. Chem. Eng. Data 55 (2010) 5406-5412.
[75] M.M. Mato, M. López, J.L Legido, J. Salgado, P.V. Verdes, M.I. Paz-Andrade, J. Chem. Eng. Data 48 (2003) 646-651.
[76] H.V. Kehaian, M.R. Tiné, L. Lepori, E. Matteoli, B. Marongiu, Fluid Phase Equilib. 46 (1989) 131-177.
[77] H.V. Kehiaian, B. Marongiu, Fluid Phase Equilib. 40 (1988) 23-78.
[78] J.A. González, Ind. Eng. Chem. Res. 50 (2011) 9810-9820.
[79] J.A. González, I. Alonso, I. García de la Fuente, J.C. Cobos, Fluid Phase Equililib. 356 (2013) 117-125.
[80] J. Wang, X. An. N. Wang, H. Lv, S. Chai, W. Shen, J. Chem. Thermodyn. 40 (2008) 1638-1644.
[81] A. Alessandrini, P. Alessi, I. Kikic, An. Chim. (Rome) 70 (1980) 293-300.
[82] I.A. McLure, A.T. Rodríguez, P.A. Ingham J.F. Steele, Fluid Phase Equilib. 8 (1982) 271-284.
[83] J.S. Chickos, W.E. Acree, J. Phys. Chem. Ref. Data 32 (2003) 519-878.
[84] F.W. Evans, H.A. Skinner, Trans. Faraday Soc. 55 (1959) 255-259.
[85] V. Majer, V. Svoboda, Enthalpies of Vaporization of Organic Compounds. Blackwell, Oxford, (1985).
[86] H.V. Kehiaian, K. Sosnkowska-Kehiaian, R. Hryniewicz, J. Chim. Phys.-Chim. Biol. 68 (1971) 922-934.
[87] J. Gmehling, J. Chem. Eng. Data 27 (1982) 371-373.
[88] A.W. Francis, Critical Solution Temperatures, Advances in Chemistry Series No 31: American Chemical Society, Washington DC, (1961).
[89] V. Alonso, I. Alonso, I. Mozo, J.A. González, I. García de la Fuente, J.C. Cobos, J. Chem. Eng. Data 55 (2010) 2263-2266.
[90] H. Yoshikawa, T. Kanahira, M. Kato, Fluid Phase Equilib. 94 (1994) 255-265.
[91] R. Martínez, J.A. González, I. García de la Fuente, J.C. Cobos, J. Chem. Eng. Data 45 (2000) 1036-1039.
[92] A. Xueqin, S. Weiguo, J. Chem. Thermodyn. 26 (1994) 461-468.
[93] C. Alonso-Tristán, J.A. González, I. García de la Fuente, J.C. Cobos, J. Chem. Eng. Data 58 (2013) 2339-2344.

[94] S. Malanowski, H.-J. Bittrich, D. Lempe, K. Reinhardt, J.-U. Wüstling, Fluid Phase Equilib. 98 (1994) 163-171.

[95] R. Alcalde, S. Aparicio, M.J. Dávila, B. García, J.M. Leal, Fluid Phase Equilib. 266 (2008) 90-100.

[96] J. Lobos, I. Mozo, M. Fernández Regúlez, J.A. González, I. García de la Fuente, J.C. Cobos, J. Chem. Eng. Data 51 (2006) 623-627.

[97] H. Matsuda, D. Taniguchi, J. Hashimoto, K. Kurihara, K. Ochi, K. Kojima, Fluid Phase Equilib. 260 (2007) 81-86.

[98] R. Eustaquio-Rincón, A. Romero-Martínez, A. Trejo, Fluid Phase Equilib. 91 (1993) 187-201.

[99] M. Aboy, S. Villa, N. Riesco, J.A. González, I. García de la Fuente, J.C. Cobos. J. Chem. Eng. Data 47 (2002) 950-953.

[100] H. Kalali, F. Kohler, P. Svejda, Fluid Phase Equilib., 20 (1985) 75-80.

[101] L. Lepori, P. Gianni, E. Matteoli, J. Solution Chem. 42 (2013) 1263-1304.

[102] F. Hevia, A. Cobos, J.A. González, I. García de la Fuente, V. Alonso, J. Solution Chem. 46 (2017) 150-174.

[103] E. Calvo-Iglesias, R. Bravo, M. Pintos, A. Amigo, A.H. Roux, G. Roux-Desgranges, J. Chem. Thermodyn. 39 (2007) 561-567.

[104] L. Lohmann, J. Gmehling, J. Chem. Eng. Data 46 (2001) 333-336.

[105] Y.P. Handa, G.C. Benson, Fluid Phase Equilib. 4 (1980) 261-268.

[106] J.R. Goates, R.J. Sullivan, J.B. Ott, J. Phys. Chem. 63 (1959) 589-594.

[107] J.B. Ott, J. Rex Goates, A.H. Budge, J. Phys. Chem. 66 (1962) 1387-1390.

[108] J.-P.E. Grolier, A. Faradjzadeh, H.V. Kehiaian, Thermochim. Acta 53 (1982) 157-162.

[109] W. Haijun, C. Mingzhi, Z. Guokang, J. Chem. Thermodyn. 27 (1995) 815-819.

[110] R.V. Mrazek, H.C. Van Ness, AIChE J. 7 (1961) 190-195.

[111] A.N. Prajapati, A.D. Vyas, V.A. Rana, S.P. Bhatnagar, J. Mol. Liq. 151 (2010) 12-16.

[112] D. Decroocq, Bull. Soc. Chim. Fr. (1964) 127-136.

[113] A. Chelkowski. Dielectric Physics, Elsevier, Warsaw (1980).

[114] C. Moreau, G. Douhéret, J. Chem. Thermodyn. 8 (1978) 403-410.

[115] J.C.R. Reis, T.P. Iglesias, Phys. Chem. Chem. Phys. 13 (2011) 10670-10680.

[116] J.A. González, L.F. Sanz, I. García de la Fuente, J.C. Cobos, J. Chem. Thermodyn. 91 (2015) 267-278.

[117] Ch. V.V. Ramana, A.B.V.K. Kumar, A.S. Kumar, M.A. Kumar, M.K. Moodly, Thermochim. Acta 566 (2013) 130-136.

[118] S. Gahylan, M. Rani, I. Lee, I. Moon, S.K. Maken, Korean. J. Chem. Eng. 32 (2015) 168-177.

[119] J.A. González, I. Mozo, I. García de la Fuente, J.C. Cobos, Thermochim. Acta 441 (2006) 53-68.

[120] J.A. Riddick, W.B. Bunger and T.K. Sakano, Organic Solvents, Techniques of Chemistry, Vol. II, A, Weissberger Ed., Wiley, N.Y. (1986).

[121] www.chemicalbook.com

[122] J.L. Trenzado, J.S. Matos, R. Alcalde, Fluid Phase Equilib. 200 (2002) 295-315.

[123] U. Domanska, M. Marciniak. Fluid Phase Equilib., 251 (2007) 161-166.

[124] P.S. Nikam, B.S. Jagdale, A.B. Sawant, M. Hasan, J. Chem. Eng. Data, 45 (2000) 214-218.

[125] Y. Lei, Z. Chen. N. Wang, C. Mao, X. An, W. Shen, J. Chem. Thermodyn. 52 (2010) 864-872.

[126] T. Yin, Y. Lei, C. Mao, Z. Chen, X. An, W. Shen, J. Chem. Thermodyn. 53 (2012) 42-51.

[127] C. Mao, N. Wang, X. Peng, X. An, W. Shen, J. Chem. Thermodyn. 40 (2008) 424-430.

[128] T. Yin, Y. Lei, M. Huang, Z. Chen, C. Mao, X. An, W. Shen, J. Chem. Thermodyn. 43 (2011) 656-663.

[129] Z. Chen, Y. Bai, T. Yin, X. An, W. Shen, J. Chem. Thermodyn. 54 (2012) 438-443.

[130] N. Wang, C. Mao, X. Peng, X. An, W. Shen, J. Chem. Thermodyn. 38 (2006) 732-738.

[131] N. Wang, C. Mao, R. Lu, X. Peng, X. An, W. Shen, J. Chem. Thermodyn. 38 (2006) 264-271.

TABLE 1

Properties of pure compounds at 0.1 MPa and 298.15 K[a]

| Compound | CAS | Source | Initial mole fraction | $\rho$ [a]/kg·m$^{-3}$ | | water[b] content |
|---|---|---|---|---|---|---|
| | | | | Exp. | Lit. | |
| Phenylacetonitrile | 140-29-4 | Sigma-Aldrich | ≥ 0.98 | 1.01289 | 1.0123[c] 1.0125[d] | 170 |
| 3-phenylpropionitrile | 645-59-0 | Sigma-Aldrich | ≥ 0.99 | 0.99644 | 1.001[e] | 104 |
| Heptane | 142-82-5 | Fluka | ≥ 0.995 | 0.67971 | 0.67946[d] | 42 |
| Octane | 111-65-9 | Sigma-Aldrich | ≥ 0.99 | 0.69871 | 0.69862[d] | 66 |
| Nonane | 111-84-2 | Fluka | ≥ 0.99 | 0.71393 | 0.71375[d] | 26 |
| Cyclooctane | 292-64-8 | Fluka | ≥ 0.99 | 0.83161 | 0.83177[f] | 14 |
| 2,2,4-trimethylpentane | 59045 | Fluka | ≥ 0.995 | 0.68732 | 0.68781[e] | 28 |

[a]standard uncertainties are: $u(T)=0.02$ K; $u(p)=1$ kPa; relative standard uncertainties are: $u_r(\rho)$ = 0.0012 (density) and 0.02 for water content; [b]in mass fraction; [c][109]; [d][120]; [e][121]; [f][122]

TABLE 2

Experimental liquid-liquid equilibrium temperatures for aromatic nitrile(1) + alkane(2) mixtures[a] at 0.1 MPa.

| $x_1$ | $T$/K | $x_1$ | $T$/K |
|---|---|---|---|
| $C_6H_5$-$CH_2CN$(1) + heptane(2) | | | |
| 0.1573 | 333.9 | 0.3979 | 350.2 |
| 0.1941 | 344.3 | 0.4251 | 350.1 |
| 0.2051 | 346.0 | 0.4704 | 350.2 |
| 0.2103 | 346.7 | 0.5025 | 349.9 |
| 0.2372 | 348.4 | 0.5505 | 349.2 |
| 0.2547 | 349.4 | 0.5996 | 347.7 |
| 0.2709 | 349.7 | 0.6478 | 345.3 |
| 0.2942 | 350.1 | 0.6702 | 343.7 |
| 0.3206 | 350.2 | 0.7169 | 338.2 |
| 0.3454 | 350.3 | 0.7365 | 336.2 |
| 0.3791 | 350.2 | 0.7517 | 333.4 |
| $C_6H_5$-$CH_2CN$(1) + octane(2) | | | |
| 0.1987 | 348.5 | 0.5039 | 352.2 |
| 0.2424 | 350.9 | 0.5293 | 352.3 |
| 0.2866 | 352.5 | 0.5447 | 352.1 |
| 0.3005 | 352.6 | 0.5923 | 351.1 |
| 0.3203 | 352.7 | 0.6240 | 350.6 |
| 0.3568 | 352.8 | 0.6711 | 348.7 |
| 0.3723 | 352.8 | 0.7021 | 346.1 |
| 0.4115 | 352.7 | 0.7430 | 340.0 |
| 0.4587 | 352.5 | 0.7646 | 336.2 |
| 0.4883 | 352.4 | | |
| $C_6H_5$-$CH_2CN$(1) + nonane(2) | | | |
| 0.2247 | 349.4 | 0.4471 | 356.1 |
| 0.2648 | 354.0 | 0.4780 | 356.0 |
| 0.2720 | 354.9 | 0.5058 | 355.8 |
| 0.3023 | 355.9 | 0.5298 | 355.6 |
| 0.3033 | 356.0 | 0.5575 | 355.4 |
| 0.3369 | 356.0 | 0.5745 | 355.3 |
| 0.3558 | 356.2 | 0.5944 | 355.1 |
| 0.3797 | 356.2 | 0.6250 | 354.7 |
| 0.4011 | 356.2 | 0.7246 | 348.1 |

TABLE 2 (continued)

| | | | |
|---|---|---|---|
| 0.4321 | 356.2 | 0.7531 | 344.8 |
| | $C_6H_5$-$CH_2CN$(1) + cyclooctane(2) | | |
| 0.1402 | 304.4 | 0.4061 | 310.3 |
| 0.1511 | 306.6 | 0.4276 | 310.2 |
| 0.1785 | 309.1 | 0.4529 | 310.1 |
| 0.1933 | 309.5 | 0.4788 | 310.0 |
| 0.2343 | 310.2 | 0.5009 | 309.7 |
| 0.2506 | 310.5 | 0.5477 | 308.7 |
| 0.2840 | 310.6 | 0.5747 | 307.9 |
| 0.2934 | 310.6 | 0.5944 | 307.1 |
| 0.3049 | 310.6 | 0.6143 | 306.0 |
| 0.3155 | 310.6 | 0.6454 | 303.9 |
| 0.3516 | 310.5 | 0.6678 | 302.7 |
| 0.3778 | 310.4 | 0.7315 | 294.9 |
| | $C_6H_5$-$CH_2CN$(1) + 2,2,4-trimethylpentane(2) | | |
| 0.2064 | 355.6 | 0.5123 | 359.5 |
| 0.2317 | 358.0 | 0.5536 | 359.3 |
| 0.2613 | 359.3 | 0.5695 | 358.9 |
| 0.2791 | 359.7 | 0.6123 | 358.2 |
| 0.3077 | 359.9 | 0.6221 | 357.9 |
| 0.3321 | 360.3 | 0.6428 | 356.8 |
| 0.3503 | 360.3 | 0.6711 | 355.5 |
| 0.3660 | 360.4 | 0.6995 | 353.4 |
| 0.4156 | 360.2 | 0.7126 | 352.1 |
| 0.4448 | 360.1 | 0.7443 | 348.0 |
| 0.4771 | 359.9 | | |
| | $C_6H_5$-$CH_2CH_2CN$(1) + heptane(2) | | |
| 0.1756 | 341.3 | 0.4260 | 353.3 |
| 0.2040 | 346.0 | 0.4437 | 353.2 |
| 0.2355 | 349.5 | 0.4812 | 353.1 |
| 0.2444 | 350.1 | 0.5022 | 353.1 |
| 0.2505 | 350.5 | 0.5211 | 352.7 |
| 0.2745 | 351.9 | 0.5448 | 352.1 |
| 0.3066 | 352.8 | 0.5748 | 351.2 |

TABLE 2 (continued)

| | | | |
|---|---|---|---|
| 0.3555 | 353.2 | 0.5965 | 349.8 |
| 0.3547 | 353.1 | 0.6267 | 347.5 |
| 0.3763 | 353.2 | 0.6657 | 342.0 |
| 0.4023 | 353.3 | | |
| | $C_6H_5$-$CH_2CH_2CN$(1) + octane(2) | | |
| 0.1993 | 347.6 | 0.4524 | 356.4 |
| 0.2167 | 349.2 | 0.4737 | 356.5 |
| 0.2279 | 350.5 | 0.5177 | 356.4 |
| 0.2598 | 352.6 | 0.5478 | 356.3 |
| 0.2829 | 353.9 | 0.5891 | 355.6 |
| 0.2834 | 353.9 | 0.6038 | 355.1 |
| 0.3128 | 354.9 | 0.6767 | 351.9 |
| 0.3415 | 355.6 | 0.7010 | 349.5 |
| 0.3772 | 356.1 | 0.7278 | 346.0 |
| 0.4105 | 356.4 | 0.7336 | 344.5 |
| 0.4283 | 356.3 | | |

[a]standard uncertainties are: $u(x_1)$ = 0.0005; $u(p)$ = 1 kPa; the combined expanded uncertainty (0.95 level of confidence) for temperature is $U_c(T) =$ 0.2 K

TABLE 3

Coefficients in eq. (1) for the fitting of the ($x_1$, $T$) pairs given in Table 2 for aromatic nitrile (1) + alkane(2) mixtures; $\sigma(T)$ is the standard deviation defined by eq. (5).

| $N$ [a] | $M$ | $k$ | $\alpha$ | $T_c$/K | $x_{1c}$ | $\sigma(T)$/K |
|---|---|---|---|---|---|---|
| | | | $C_6H_5$-$CH_2CN$(1) + heptane(2) | | | |
| 22 | 4.699 | − 5778 | 2.293 | 350.1 | 0.3819 | 0.17 |
| | | | $C_6H_5$-$CH_2CN$(1) + octane(2) | | | |
| 19 | 4.300 | − 1930 | 1.100 | 352.6 | 0.4282 | 0.34 |
| | | | $C_6H_5$-$CH_2CN$(1) + nonane(2) | | | |
| 20 | 4.370 | − 4590 | 2.287 | 356.2 | 0.4191 | 0.27 |
| | | | $C_6H_5$-$CH_2CN$(1) + cyclooctane(2) | | | |
| 24 | 5.197 | − 6527 | 2.785 | 310.4 | 0.3240 | 0.20 |
| | | | $C_6H_5$-$CH_2CN$(1) + 2,2,4-trimethylpentane(2) | | | |
| 21 | 4.697 | − 4516 | 2.025 | 360.0 | 0.3989 | 0.27 |
| | | | $C_6H_5$-$CH_2CH_2CN$(1) + heptane(2) | | | |
| 21 | 3.858 | − 2433 | 1.153 | 353.2 | 0.4141 | 0.11 |
| | | | $C_6H_5$-$CH_2CH_2CN$(1) + octane(2) | | | |
| 21 | 3.594 | − 1257 | 0.807 | 356.3 | 0.4713 | 0.18 |

[a] number of experimental data points

TABLE 4

Dispersive (DIS) and quasichemical (QUAC) interchange coefficients, $C^{\mathrm{DIS}}_{\mathrm{sn},l}$ and $C^{\mathrm{QUAC}}_{\mathrm{sn},l}$ ($l$ = 1, Gibbs energy; $l$ = 2, enthalpy; $l$ = 3, heat capacity) for (s,n) contacts[a] in aromatic nitrile + organic solvent mixtures.

| Solvent | Contact (s,n)[a] | $C^{\mathrm{DIS}}_{\mathrm{sn},1}$ | $C^{\mathrm{DIS}}_{\mathrm{sn},2}$ | $C^{\mathrm{DIS}}_{\mathrm{sn},3}$ | $C^{\mathrm{QUAC}}_{\mathrm{sn},1}$ | $C^{\mathrm{QUAC}}_{\mathrm{sn},2}$ | $C^{\mathrm{QUAC}}_{\mathrm{sn},3}$ |
|---|---|---|---|---|---|---|---|
| | | | Benzonitrile | | | | |
| Aromatic hydrocarbon[b] | (b,n) | 1 | 0.22 | − 6.2 | 0.75 | 0.04 | 4.5 |
| Toluene | (a,n) | − 1.68 | − 2.19 | − 4.3 | 4.75 | 5 | 6 |
| Ethylbenzene | (a,n) | − 1.4 | − 1.74 | − 4.3 | 4.75 | 5 | 6 |
| Octane | (a,n) | − 1 | − 0.93 | − 4.3 | 4.75 | 5 | 6 |
| Nonane | (a,n) | − 1.09 | − 0.93 | − 4 | 4.75 | 5 | 6 |
| Undecane | (a,n) | − 1.095 | − 0.93 | − 3.5 | 4.75 | 5 | 6 |
| ≥ Dodecane | (a,n) | − 1.11 | − 0.93 | − 3.5 | 4.75 | 5 | 6 |
| Cycloalkane | (c,n) | − 1 | − 0.93 | − 4.5 | 4.75 | 5 | 6 |
| Methanol | (h,n) | 5[c] | − 12.8 | | − 0.75 | 8.2 | |
| Ethanol | (h,n) | 5[c] | − 12.8 | | − 0.75 | 7.9 | |
| ≥ 1-propanol | (h,n) | 1 | 3.1 | | 6 | − 0.2 | |
| | | | Phenylacetonitrile | | | | |
| Aromatic hydrocarbon[b] | (b,n) | 1 | 0.22 | − 6.2 | 0.75 | 0.04 | 4.5 |
| Heptane | (a,n) | 0.08 | − 0.4[c] | − 4.3 | 4.75 | 5 | 6 |
| Octane | (a,n) | − 0.025 | − 0.4[c] | − 4.3 | 4.75 | 5 | 6 |
| Nonane | (a,n) | − 0.09 | − 0.4[c] | − 4 | 4.75 | 5 | 6 |
| Cycloalkane | (c,n) | − 0.025 | − 0.4[c] | − 4.3 | 4.75 | 5 | 6 |
| | | | 3-Phenylpropionitrile | | | | |
| Aromatic hydrocarbon[b] | (b,n) | 1 | 0.22 | − 6.2 | 0.75 | 0.04 | 4.5 |
| Heptane | (a,n) | 0.605 | − 0.4[c] | − 4.3 | 4.75 | 5 | 6 |
| Octane | (a,n) | 0.46 | − 0.4[c] | − 4.3 | 4.75 | 5 | 6 |

[a]type s = a, $CH_2$, $CH_3$ in *n*-alkanes, 1-alkanols or aromatic hydrocarbons; s = b, $C_6H_6$ or $C_6H_5-$ in aromatic hydrocarbons or nitriles, s = c, c-$CH_2$ in cycloalkanes; s = h, OH in 1-alkanols; type n, CN in aromatic nitriles; [b]$C^{\mathrm{DIS}}_{\mathrm{bn},3} = -8$ in ethylbenzene systems; [c]guessed value

TABLE 5

ERAS parameters[a] for 1-alkanol + benzonitrile mixtures at 298.15 K and 0.1 MPa.

| 1-alkanol | $K_{AB}$ | $\Delta v^*_{AB}$ / cm$^3$·mol$^{-1}$ | $\Delta h^*_{AB}$ / kJ·mol$^{-1}$ | $X_{AB}$ / J·cm$^{-3}$ |
|---|---|---|---|---|
| Methanol | 32 | – 7.5 | – 12 | 10 |
| Ethanol | 20 | – 7.5 | – 12 | 14 |
| 1-propanol | 20 | – 7.5 | – 12 | 17 |
| 1-butanol | 18.5[b] | – 7.5 | – 12 | 17[b] |
| 1-pentanol | 18.5[b] | – 7.8 | – 12 | 17[b] |

[a] $K_{AB}$, association constant of component A with component B; $\Delta h^*_{AB}$, association enthalpy of component A with component B; $\Delta v^*_{AB}$, association volume of component A with component B; $X_{AB}$, physical parameter; [b]$T$ = 303.15 K

TABLE 6

Molar excess Gibbs energies, $G^E_m$, at equimolar composition and at temperature $T$, for benzonitrile(1) + aromatic hydrocarbon(2) mixtures [24].

| $T$/K | $N^a$ | $G^E_m$ /J·mol$^{-1}$ | | $\sigma_r(P)$ [b] | |
|---|---|---|---|---|---|
| | | Exp[c] | DQ[d] | Exp.[c] | DQ[d] |
| Benzene | | | | | |
| 323.15 | 52 | 205 | 215 | 0.008 | 0.011 |
| 353.15 | 51 | 247 | 234 | 0.008 | 0.011 |
| Toluene | | | | | |
| 323.15 | 48 | 255 | 255 | 0.004 | 0.013 |
| 353.15 | 51 | 268 | 276 | 0.003 | 0.012 |

[a]number of data points; [b]equation (12); [c]experimental value; [d]DISQUAC value calculated with interaction parameters from Table 4

TABLE 7

SLE results from DISQUAC (DQ), using interaction parameters from Table 4, and from the Ideal Solubility Model for benzonitrile + organic solvent mixtures.

| | $N^a$ | $\Delta(T)$ [b]/K | | $\sigma_r(T)$ [c] | | Ref. |
|---|---|---|---|---|---|---|
| | | Ideal | DQ | Ideal | DQ | |
| Benzene | 19 | 1.5 | 0.40 | 0.008 | 0.002 | 104 |
| Toluene | 15 | 1.3 | 1.4 | 0.007 | 0.009 | 104 |
| 1-octanol | 29 | 6.1 | 1.1 | 0.029 | 0.005 | 123 |
| 1-decanol | 23 | 8.4 | 2.0 | 0.039 | 0.009 | 123 |

[a]number of data points; [b]absolute mean deviation, $\Delta(T)/K = \frac{1}{N}\sum\left|T_{\text{exp}} - T_{\text{calc}}\right|$; [c] standard relative deviation (eq. 12)

TABLE 8

Coordinates of the critical points of aromatic nitrile(1) + alkane(2) mixtures

| Alkane | $x_{1c}$ | | $T_c$ /K | | Ref. |
|---|---|---|---|---|---|
| | Exp.[a] | DQ[b] | Exp.[a] | DQ[b] | |
| Benzonitrile | | | | | |
| Octane | 0.464 | 0.452 | 283.2 | 283.6 | 80 |
| Nonane | 0.497 | 0.492 | 284.6 | 286.2 | 125 |
| Undecane | 0.441 | 0.562 | 290.1 | 291.8 | 126 |
| Dodecane | 0.586 | 0.594 | 293.1 | 293.4 | 127 |
| Tetradecane | 0.633 | 0.663 | 298.9 | 300.6 | 128 |
| Pentadecane | 0.657 | 0.682 | 301.2 | 303.7 | 129 |
| Hexadecane | 0.673 | 0.706 | 304.4 | 306.7 | 130 |
| Heptadecane | 0.695 | 0.728 | 306.6 | 309.5 | 129 |
| Octadecane | 0.707 | 0.739 | 309.6 | 312.2 | 131 |
| Phenylacetonitrile | | | | | |
| Heptane | 0.382 | 0.468 | 350.1 | 351.2 | This work |
| Octane | 0.428 | 0.507 | 352.6 | 355.2 | This work |
| Nonane | 0.419 | 0.537 | 356.2 | 359.1 | This work |
| Cyclooctane | 0.324 | 0.329 | 310.4 | 310.6 | This work |
| 3-Phenylpropinitrile | | | | | |
| Heptane | 0.414 | 0.427 | 353.2 | 354.7 | This work |
| Octane | 0.469 | 0.473 | 356.3 | 358.7 | This work |

[a]experimental value; [b]DISQUAC value calculated with interaction parameters from Table 4

TABLE 9

Molar excess enthalpies, $H_m^E$ , at equimolar composition and at temperature *T*, and 0.1 MPa for benzonitrile(1) + organic solvent(2) mixtures.

| Solvent | *T*/K | *N*[a] | $H_m^E$/J·mol$^{-1}$ | | $dev(H_m^E)$ [b] | | Ref. |
|---|---|---|---|---|---|---|---|
| | | | Exp[c]. | DQ[d] | Exp[c]. | DQ[d] | |
| Benzene | 293.15 | 19 | 32 | 34 | 0.003 | 0.232 | 23 |
| | 298.15 | 19 | 32 | 32 | 0.006 | 0.234 | 23 |
| | | 12 | 31 | | 0.042 | 0.200 | 25 |
| | | 21 | 35 | | 0.004 | 0.243 | 52 |
| | 323.15 | 15 | 23 | 18 | 0.113 | 0.404 | 24 |
| | 363.15 | 15 | 9.8 | – 5 | 0.102 | 1.43 | 24 |
| | 413.15 | 21 | – 20 | – 35 | 0.095 | 0.850 | 24 |
| Toluene | 293.15 | 19 | – 15 | – 22 | 0.007 | 0.733 | 23 |
| | 298.15 | 19 | – 12 | – 17 | 0.008 | 0.833 | 23 |
| | | 11 | – 9 | | 0.042 | 0.947 | 25 |
| | | 16 | – 13 | | 0.015 | 0.669 | 52 |
| | 323.15 | 13 | – 11 | 12 | 0.043 | 0.600 | 24 |
| | 363.15 | 15 | 47 | 62 | 0.036 | 0.185 | 24 |
| | 413.15 | 19 | 82 | 128 | 0.012 | 0.390 | 24 |
| ethylbenzene | 298.15 | 12 | 105 | 111 | 0.014 | 0.124 | 25 |
| Cyclohexane | 298.15 | 19 | 1390 | 1385 | 0.002 | 0.007 | 52 |
| Methanol | 298.15 | 12 | 976 | 1004 | 0.009 | 0.017 | 22 |
| | | | | | | (0.022)[e] | |
| Ethanol | 298.15 | 12 | 1209 | 1206 | 0.002 | 0.099 | 22 |
| | | | | | | (0.022)[e] | |
| 1-propanol | 298.15 | 12 | 1454 | 1438 | 0.005 | 0.055 | 22 |
| | | | | | | (0.070)[e] | |

[a]number of data points; [b]equation (13); [c]experimental value; [d]DISQUAC value calculated with interaction parameters from Table 4; [e]ERAS result using parameters from Table 5

TABLE 10

Molar excess volumes, $V_m^E$ , at equimolar composition and at temperature *T*, for 1-alkanol (1) + benzonitrile(2).

| Solvent | *T*/K | $V_m^E$ /cm$^3$·mol$^{-1}$ | | Ref. |
|---|---|---|---|---|
| | | Exp[a]. | ERAS[b] | |
| Methanol | 298.15 | – 0.358 | – 0.360 | 22 |
| Ethanol | 298.15 | – 0.329 | – 0.337 | 22 |
| 1-propanol | 298.15 | – 0.262 | – 0.259 | 22 |
| 1-butanol | 303.15 | – 0.151 | – 0.157 | 124 |
| 1-pentanol | 303.15 | – 0.168 | – 0.156 | 124 |

[a]experimental value; [b]ERAS result using interaction parameters from Table 5

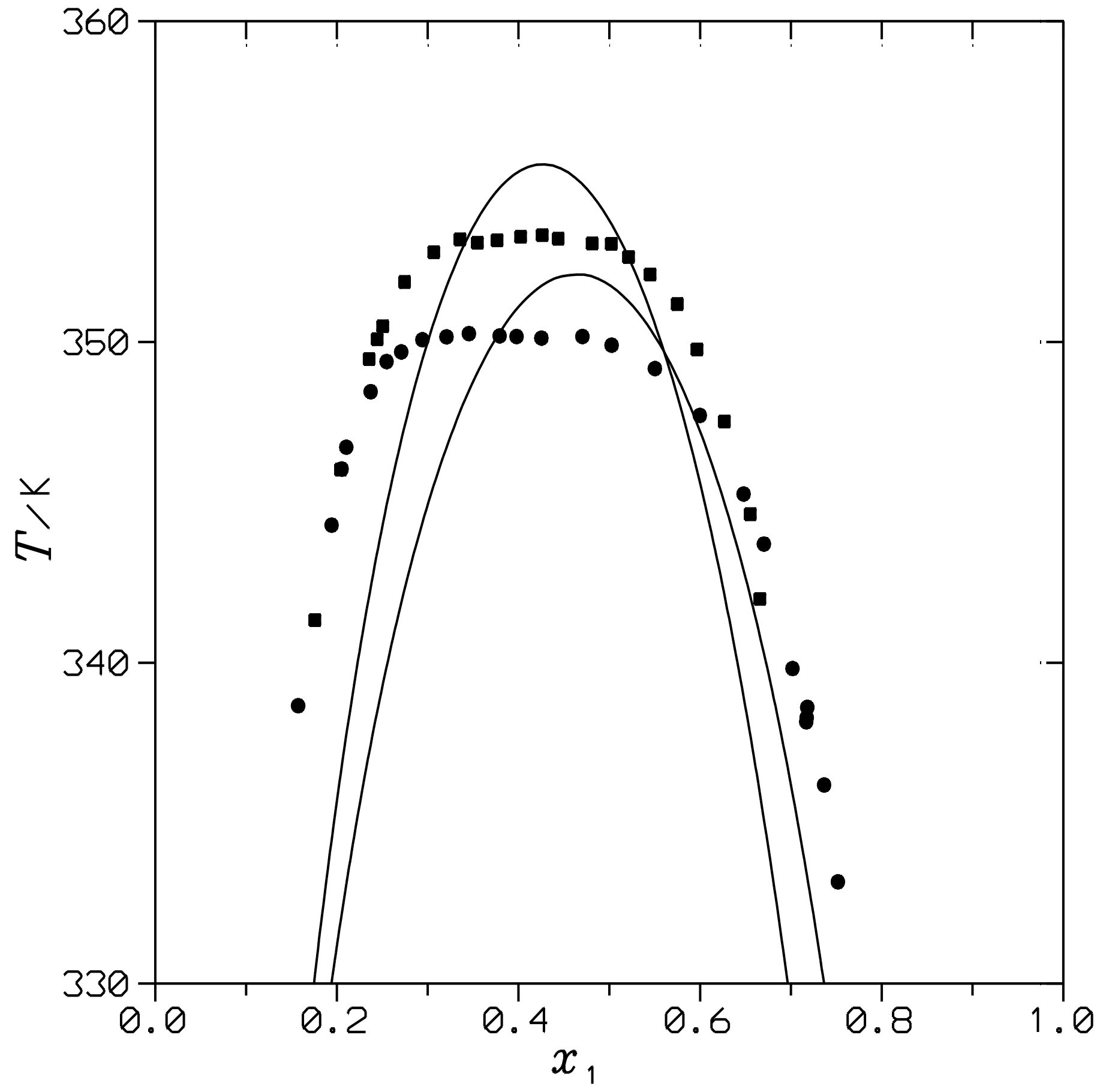

**Figure 1**

LLE for aromatic nitrile(1) + heptane(2) mixtures. Points, experimental results (this work): (●), phenylacetonitrile; (■), 3-phenylpropionitrile. Solid lines, DISQUAC calculations with interaction parameters listed in Table 4.

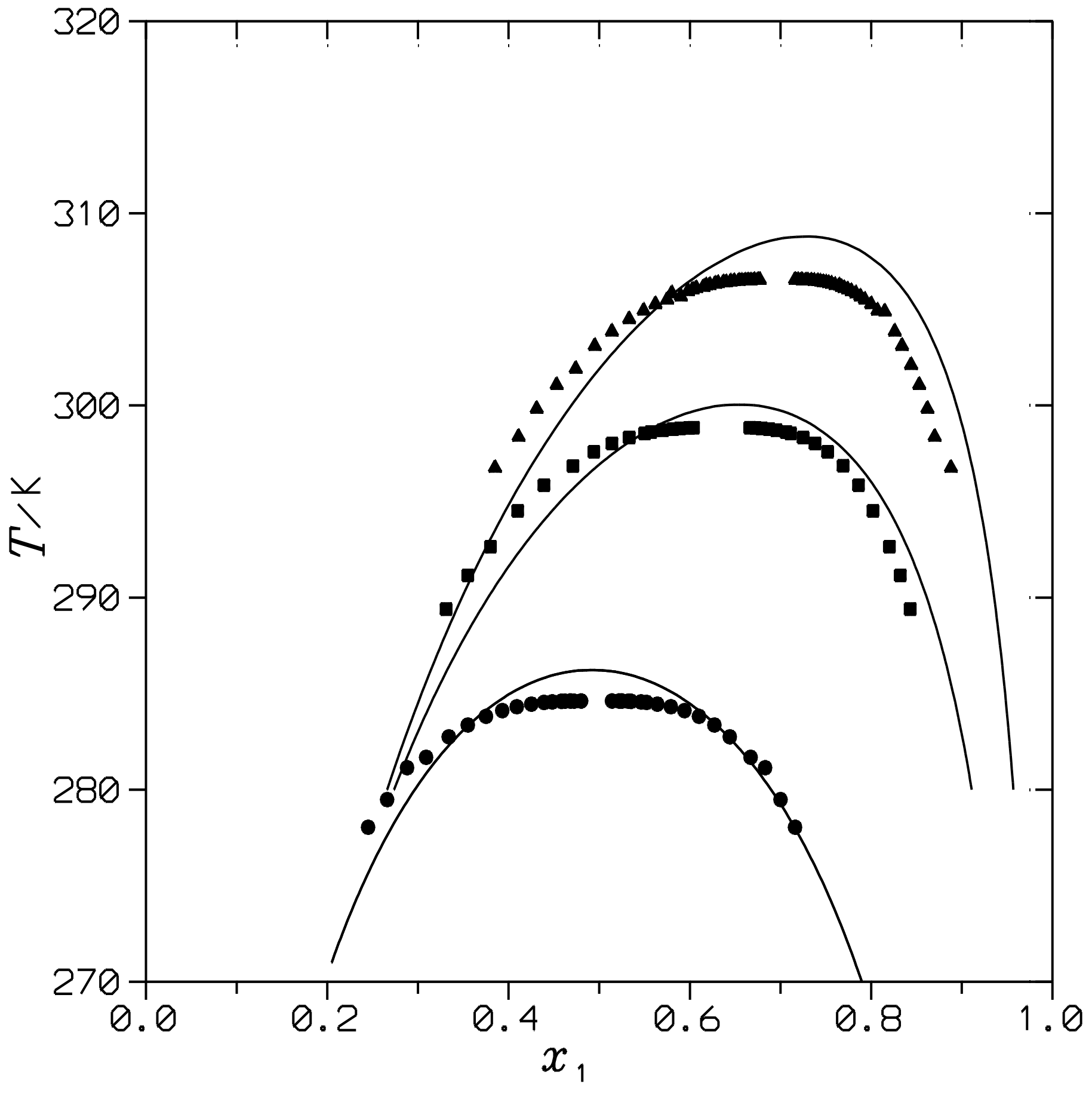

**Figure 2**

LLE for benzonitrile(1) + *n*-alkane(2) mixtures. Points, experimental results: (●), nonane [125]; (■), tetradecane [128]; (▲), heptadecane [129]. Solid lines, DISQUAC calculations with interaction parameters listed in Table 4.

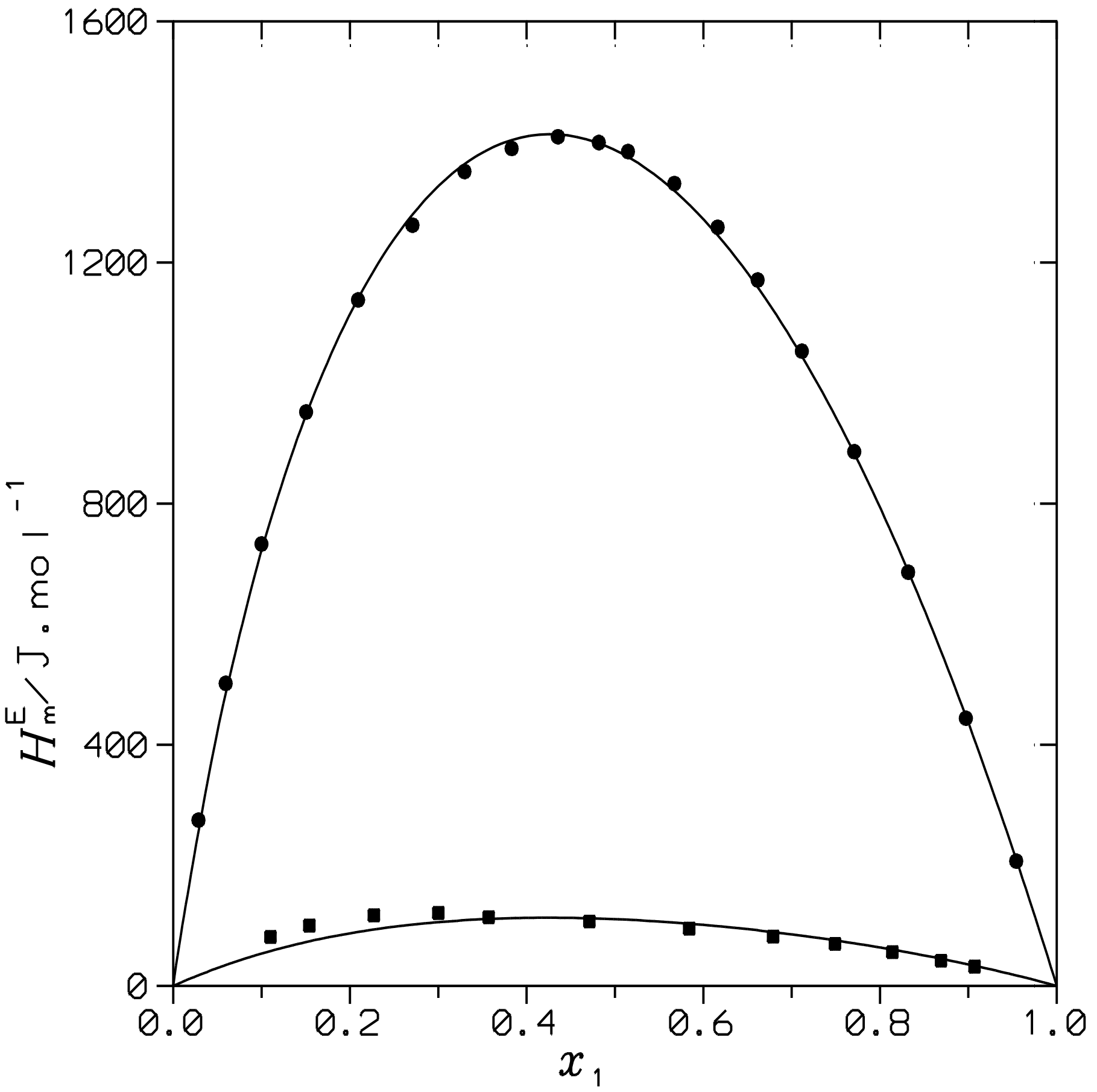

**Figure 3**

$H_m^E$ for benzonitrile(1) + cyclohexane(2) (●) [52], or + ethylbenzene(2) (■) [25] mixtures at 298.15 K. Points, experimental results. Solid lines, DISQUAC calculations with interaction parameters listed in Table 4.

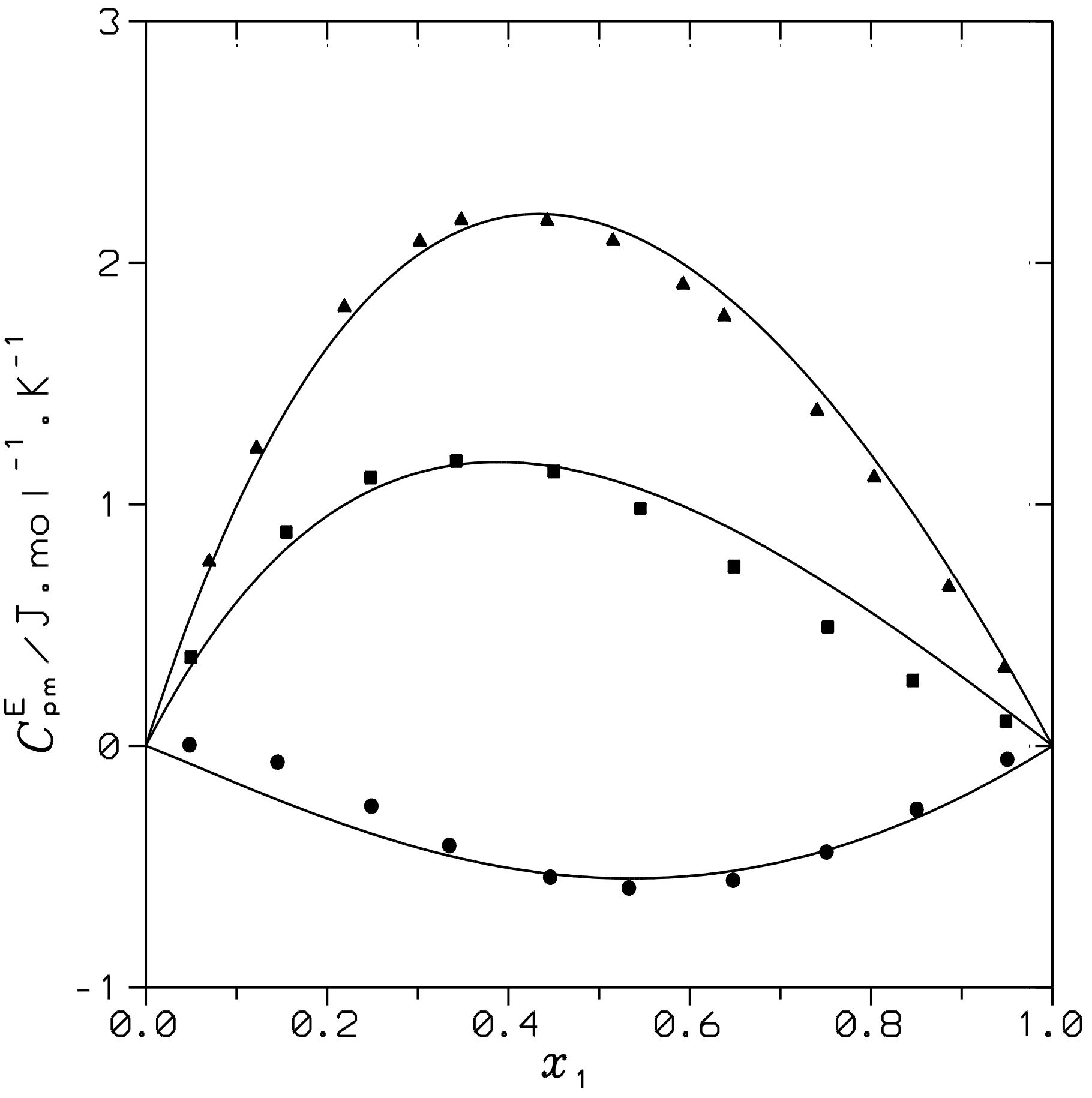

**Figure 4**

$C_{pm}^{E}$ for benzonitrile(1) + aromatic hydrocarbon(2) mixtures at 2981.5 K. Points, experimental results: (●), benzene [23]; (■), toluene [23]; (▲), ethylbenzene [103]. Solid lines, DISQUAC calculations with interaction parameters listed in Table 4.

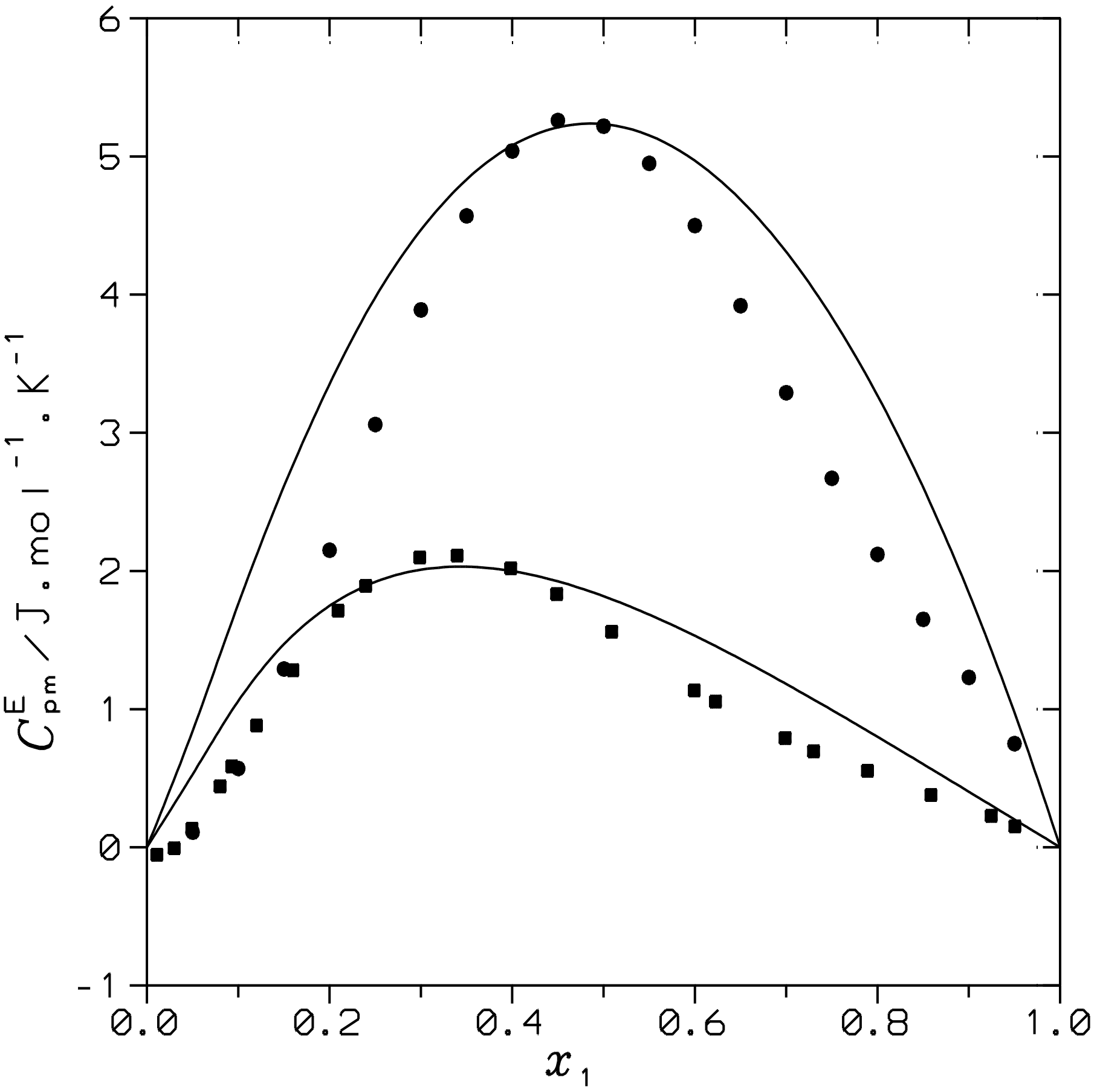

**Figure 5**

$C^{E}_{pm}$ for benzonitrile(1) + alkane(2) mixtures at 2981.5 K. Points, experimental results: (●), nonane [53]; (■). Cyclohexane [54]. Solid lines, DISQUAC calculations with interaction parameters listed in Table 4.

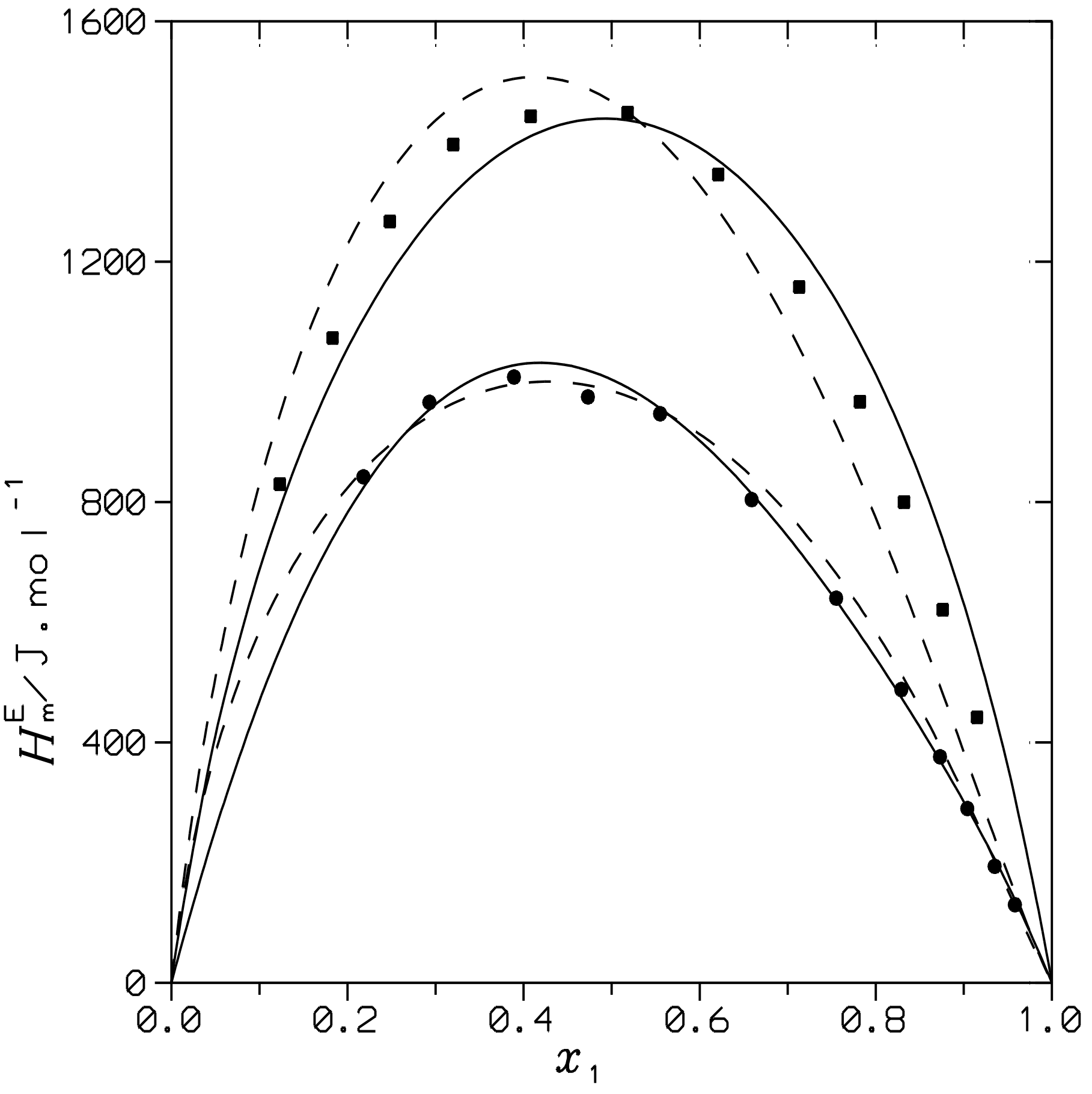

**Figure 6**

$H_m^E$ for 1-alkanol(1) + benzonitrile(2) mixtures at 298.15 K. Points, experimental results [22]: (●), methanol; (■), 1-propanol. Solid lines, DISQUAC calculations with interaction parameters listed in Table 4. Dashed lines, ERAS results using parameters from Table 5.

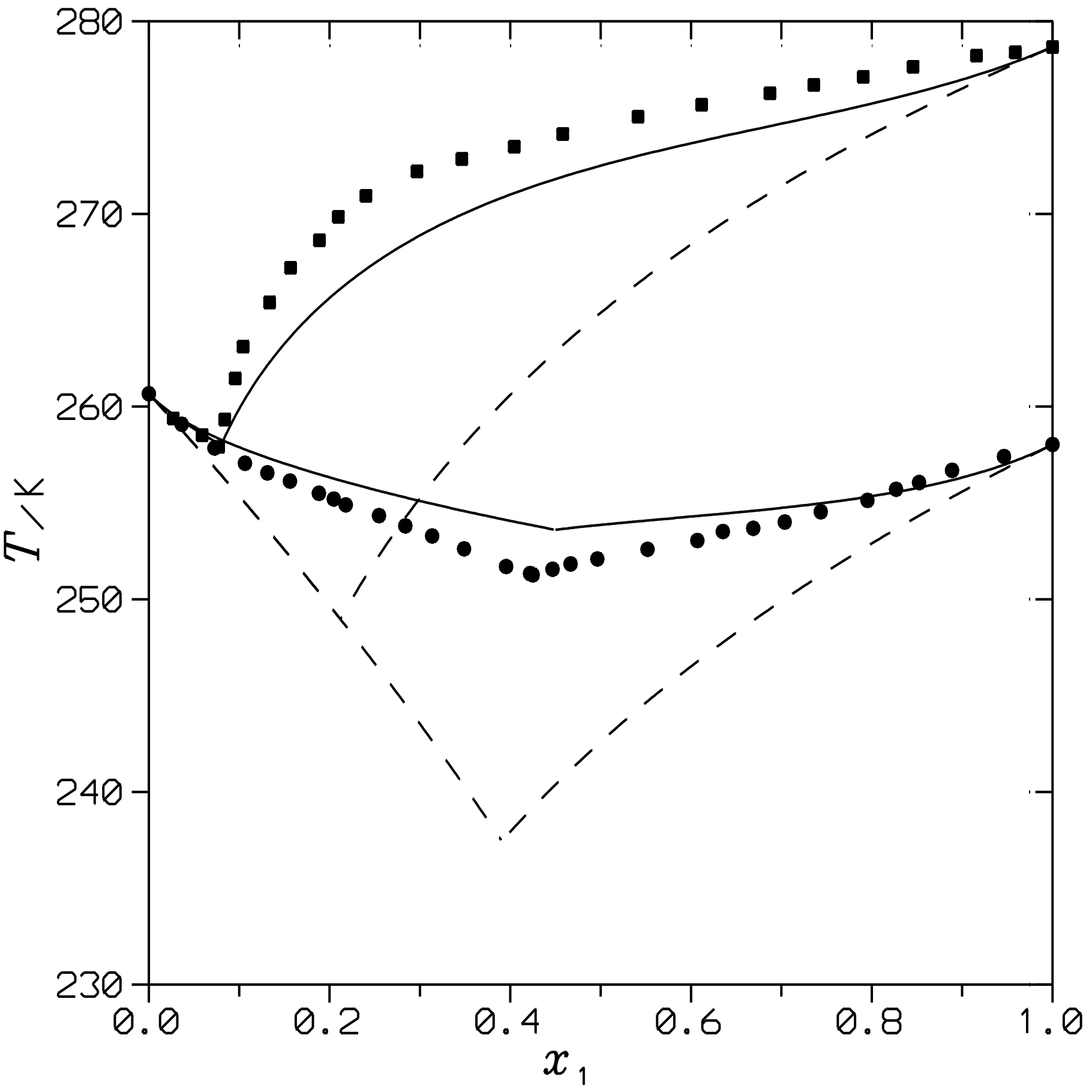

**Figure 7**

SLE for 1-alkanol(1) + benzonitrile(2) mixtures. Points, experimental results [123]: (●), 1-octanol; (■), 1-decanol. Solid lines, DISQUAC calculations with interaction parameters listed in Table 4. Dashed lines, results from the Ideal solubility model

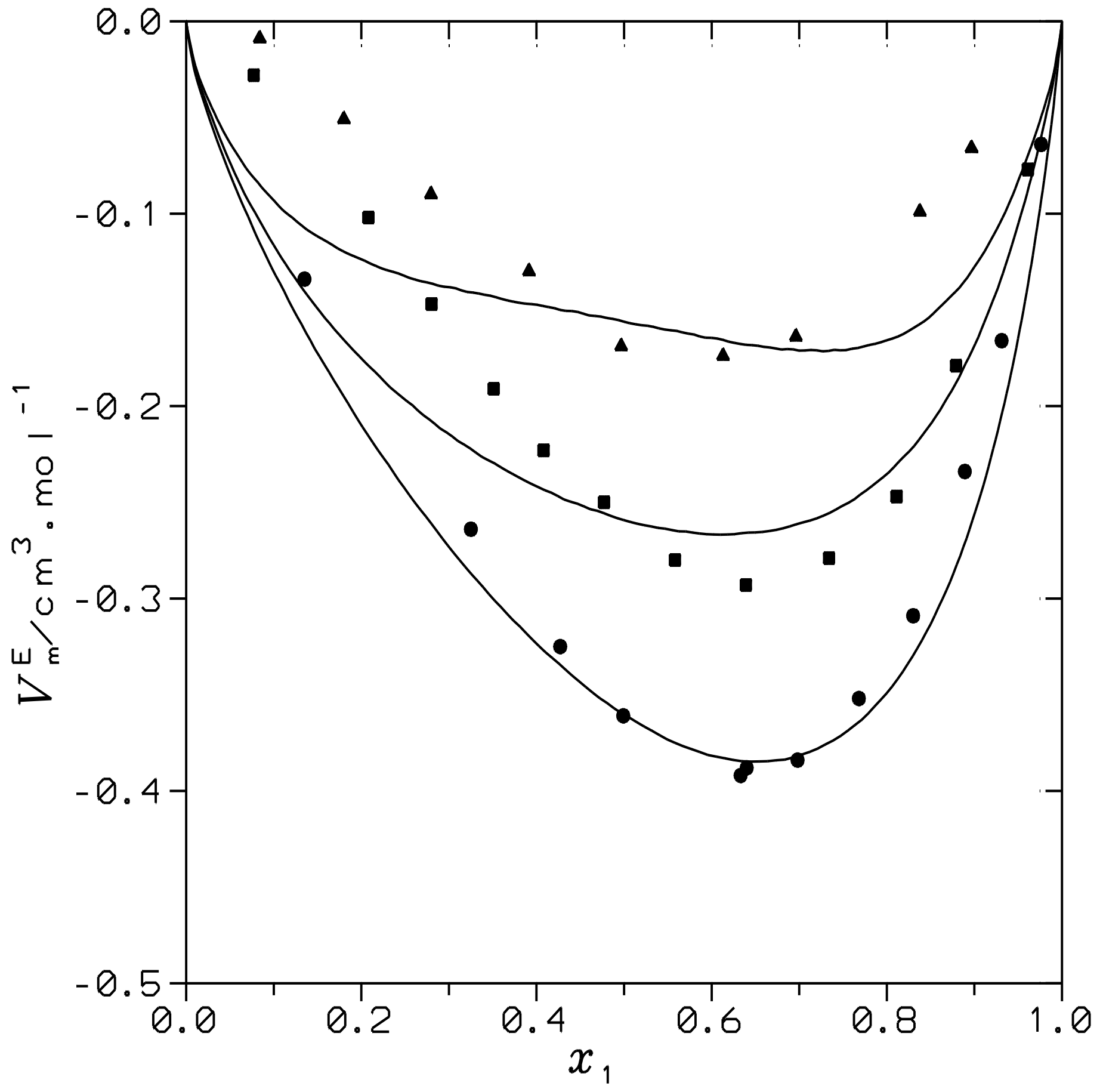

**Figure 8**

$V_m^E$ for 1-alkanol(1) + benzonitrile(2) mixtures. Points, experimental results: (●), methanol; (■), 1-propanol ($T$ = 298.15 K) [22]; (▲), 1-pentanol ($T$ = 303.15 K) [124]. Solid lines, ERAS results using parameters from Table 5.

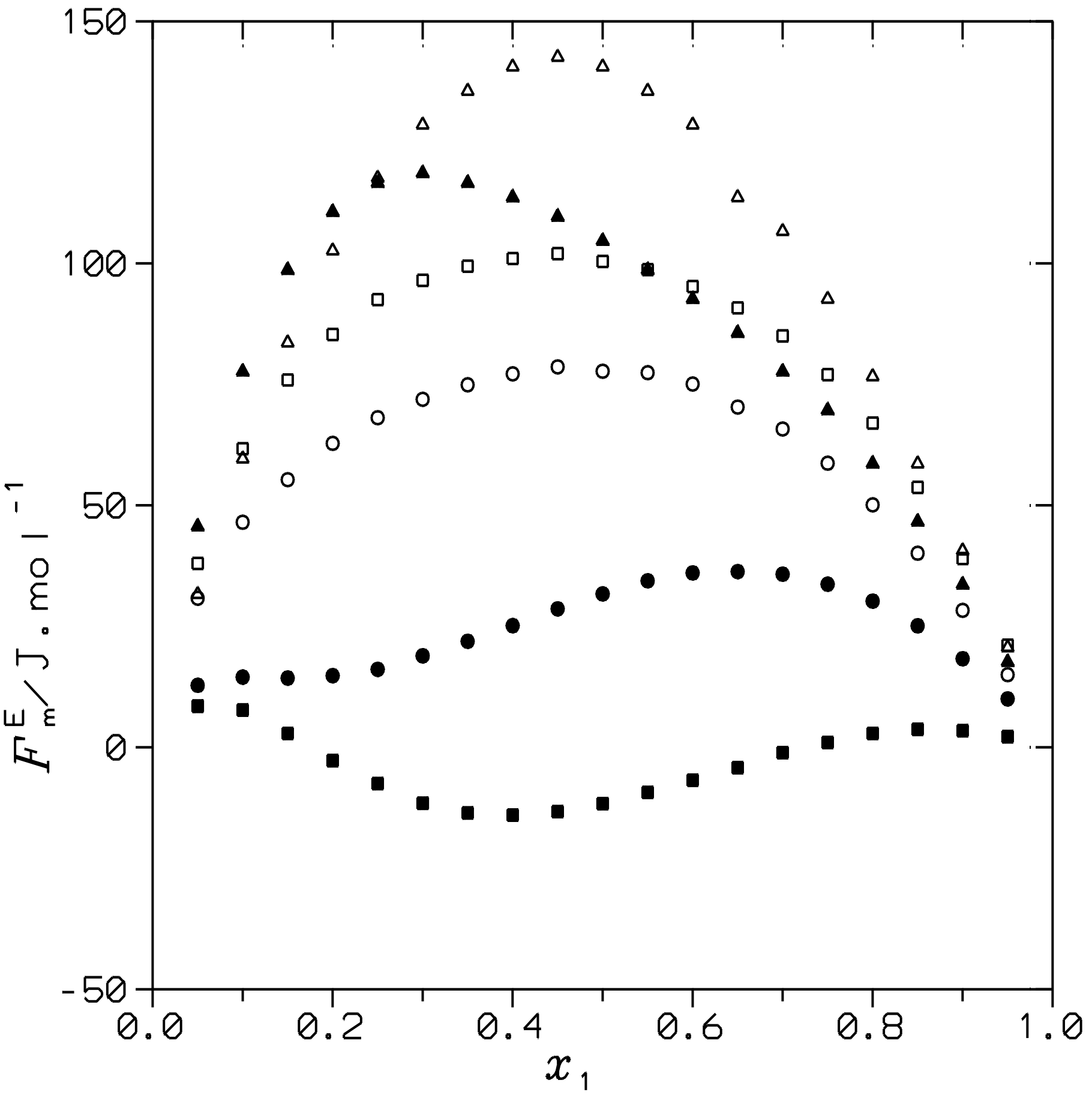

**Figure 9**

Excess molar functions for benzonitrile(1) + aromatic hydrocarbon(2) mixtures at 298.15 K. Full symbols, enthalpies ($F = H$); open symbols, isochoric internal energies ($F = U_V$) calculated using eq. (15): (●) [23], (O), benzene; (■), [23], (□), toluene; (▲) [25], (Δ), ethylbenzene.

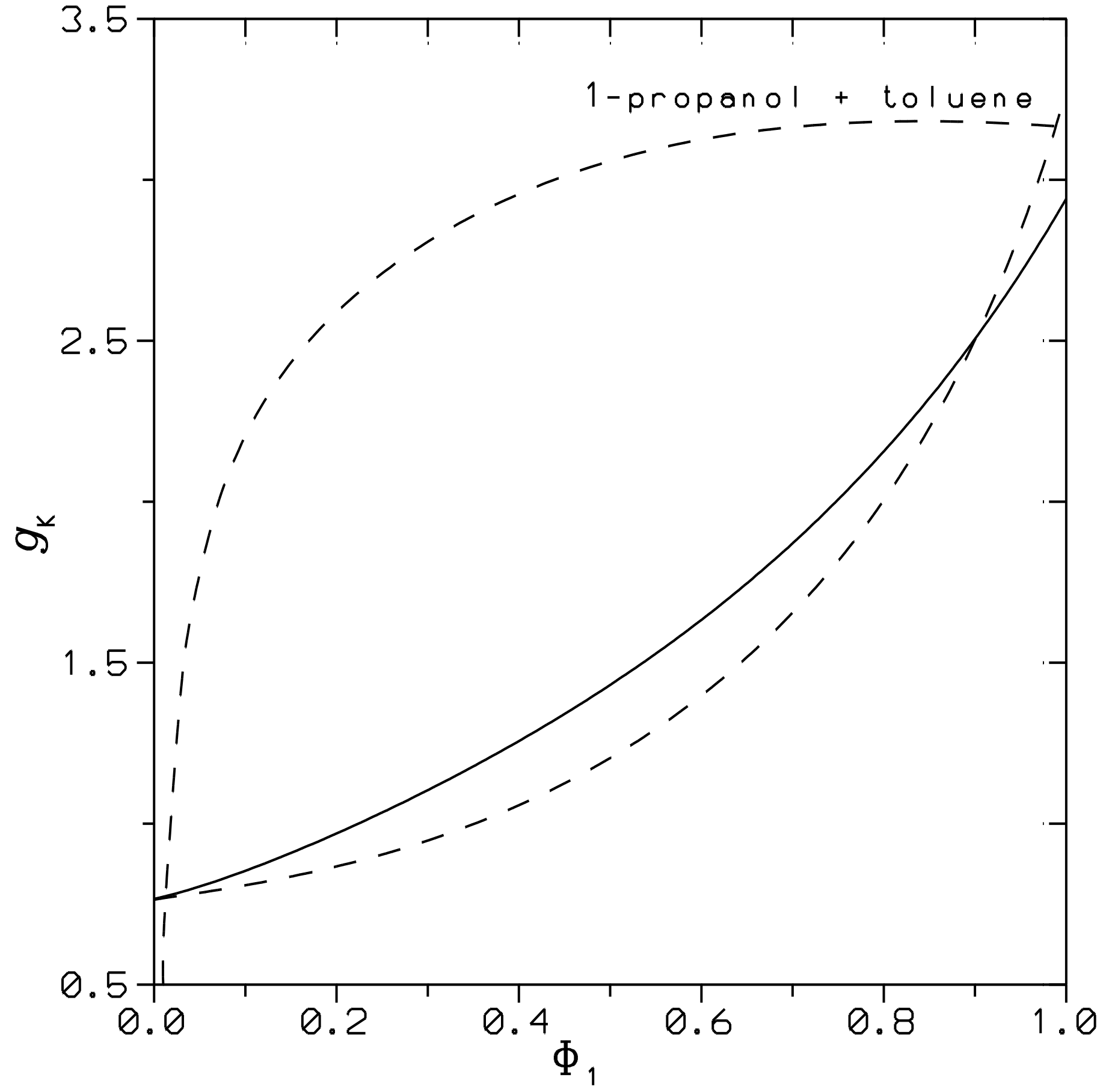

**Figure 10**

Kirkwood's correlation factor for 1-alkanol(1) + organic solvent(2) mixtures. Solid line, methanol + benzonitrile, ($T$ = 298.15 K); dashed lines, 1-propanol + benzonitrile or + toluene ($T$ = 303.15 K).

**SUPPLEMENTARY MATERIAL**

# THERMODYNAMICS OF MIXTURES CONTAINING AROMATIC NITRILES

JUAN ANTONIO GONZÁLEZ*[(1)], CRISTINA ALONSO TRISTÁN[(2)], FERNANDO HEVIA,[(1)] , ISAÍAS GARCÍA DE LA FUENTE[(1)] AND LUIS FELIPE SANZ[(1)]

[(1)] Dpto. Ingeniería Electromecánica. Escuela Politécnica Superior. Avda. Cantabria s/n. 09006 Burgos, (Spain)

[(2)] G.E.T.E.F., Departamento de Física Aplicada, Facultad de Ciencias, Universidad de Valladolid, Paseo de Belén, 7, 47011 Valladolid, Spain,

**SUPPLEMENTARY MATERIAL**

**TABLE S1**

Physical properties[a] of pure compounds at 0.1 MPa: melting temperature, $T_m$, enthalpy of fusion, $\Delta H_m$ and heat capacity change at the melting point, $\Delta C_{pm}$.

| Compound | $T_m$ /K | $\Delta H_m$ /kJ·mol$^{-1}$ | $\Delta C_{p,m}$ /J·mol$^{-1}$·K$^{-1}$ | Ref. |
|---|---|---|---|---|
| Benzonitrile | 260.67 | 10.98 | | S1 |
| | 259.75 | | | S2 |
| Benzene | 278.68 | 9.95 | | S2 |
| Toluene | 178.16 | 6.58 | | S2 |
| 1-octanol | 258.03 | 23.70 | 41.33 | S1 |
| 1-decanol | 278.67 | 31.40 | 82.65 | S1 |

**TABLE S2**

Coordinates of the eutectic points ($x_{1eu}$, $T_{eu}$) benzonitrile(1) + organic solvent(2) mixtures calculated according to the DISQUAC[a] (DQ) and ideal solubility models.

| Solvent | $x_{1eu}$ | | | $T_{eu}$ /K | | | Ref. |
|---|---|---|---|---|---|---|---|
| | Exp.[a] | DQ[b] | IDEAL[c] | Exp.[a] | DQ[b] | IDEAL[c] | |
| Benzene | 0.596 | 0.586 | 0.565 | 237.0 | 237.0 | 237.4 | S2 |
| Toluene | | 0.069 | 0.086 | | 175.5 | 174.7 | S2 |
| 1-octanol | 0.575 | 0.558 | 0.604 | 256.56 | 253.6 | 237.4 | S1 |
| 1-decanol | 0.923 | 0.920 | 0.788 | 257.91 | 258.0 | 249.0 | S1 |

[a]experimental value; [b]DISQUAC results using interaction parameters from Table 4; [c]results from the Ideal Solubility Model

**REFERENCES FOR SUPPLEMENTARY MATERIAL**

[S1] U. Domanska, M. Marciniak. Fluid Phase Equilib. 251 (2007) 161-166.

[S2] L. Lohmann, J. Gmehling, J. Chem. Eng. Data 46 (2001) 333-336.

[S3] S. Horstmann, H. Gradeler, R. Bölts, J. Gmehling, J. Chem. Eng. Data 44 (1999) 539-543.

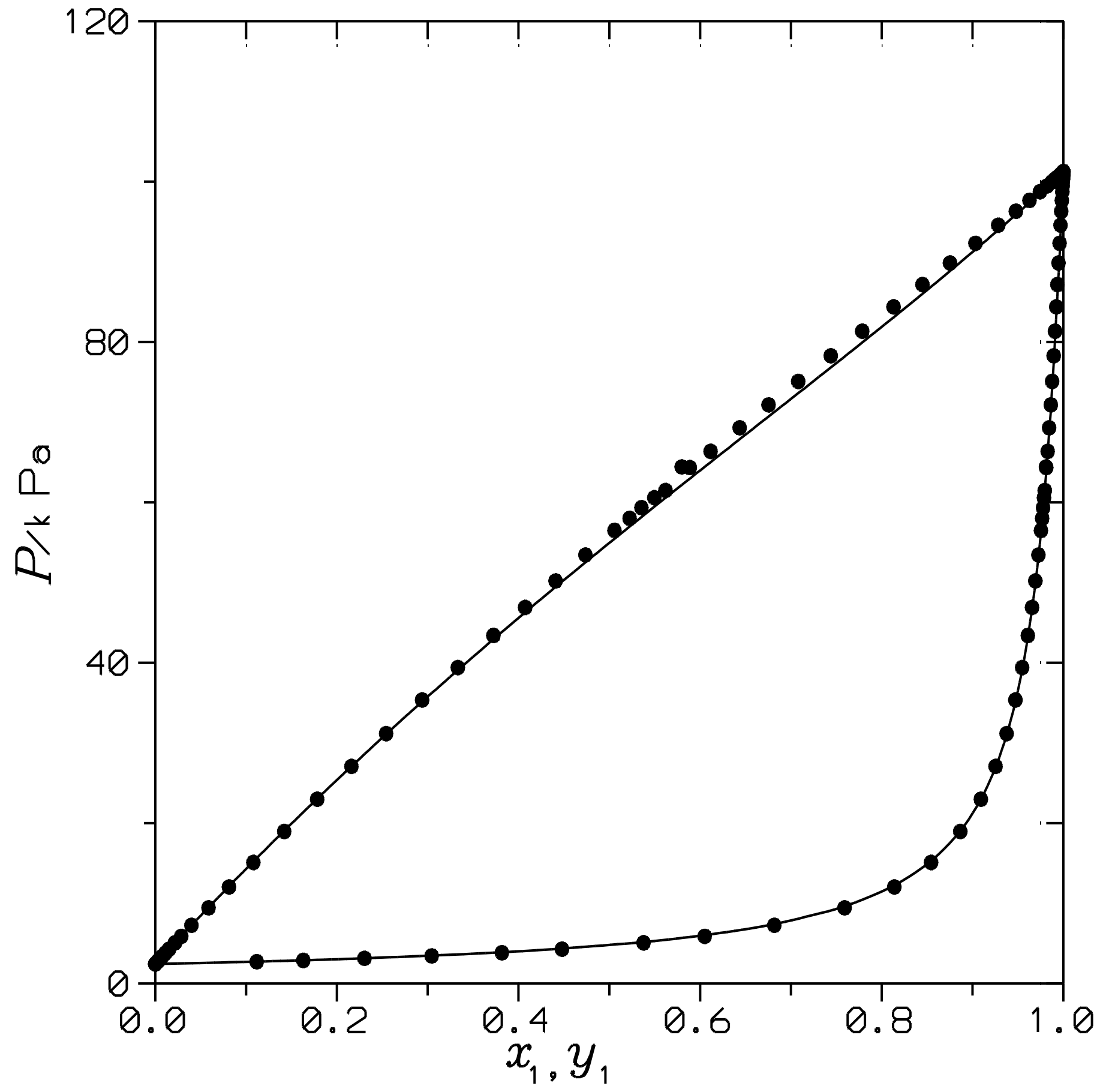

**Figure S1**

VLE for the benzene(1) + benzonitrile(2) system at 353.15 K. Points, experimental results [S3]. Solid lines, DISQUAC calculations with interaction parameters from Table 4.

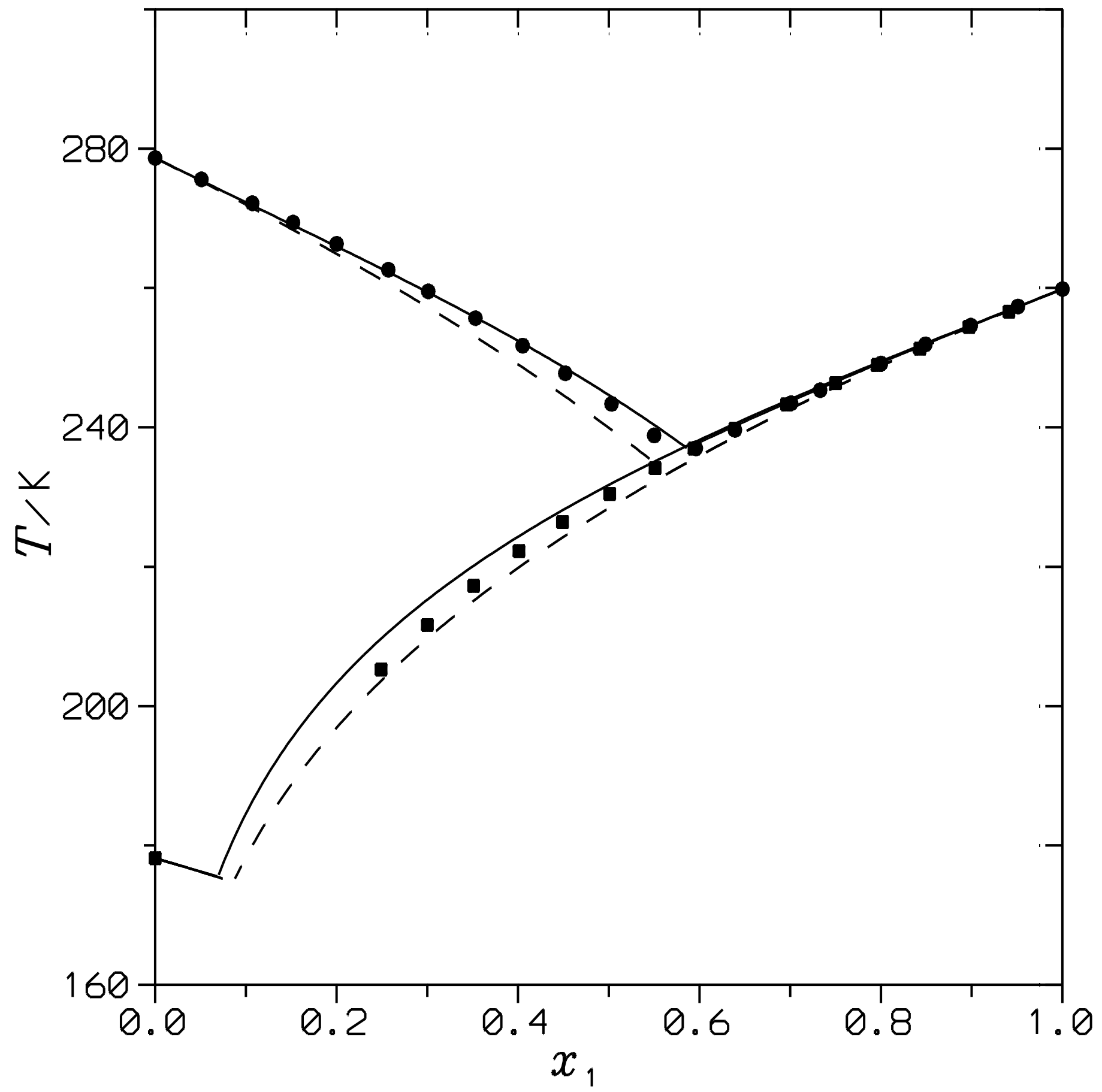

**Figure S2**

SLE for benzonitrile(1) + aromatic hydrocarbon(2) systems. Points, experimental results [S2]: (●), benzene, (●), toluene. Solid lines, DISQUAC calculations with interaction parameters from Table 4. Dashed lines, results from the ideal solubility model.